\documentclass[useAMS,usenatbib,usegraphicx]{mn2e}

\usepackage{paper_def}
\usepackage{txfonts}
\usepackage[T1]{fontenc}
\usepackage{aecompl}
\usepackage{color}
\usepackage{lscape}

\title[Widespread star formation inside galactic outflows]{Widespread star formation inside galactic outflows}
\author[Gallagher et al.]{R.~Gallagher,$^{1,2}$
R.~Maiolino,$^{1,2}$ F. Belfiore,$^{3}$
N.~Drory,$^{4}$ R. Riffel,$^{5,6}$ R.A.~Riffel,$^{6,7}$
\\
$^{1}$Cavendish Laboratory, University of Cambridge, 19 J. J. Thomson Ave., Cambridge CB3 0HE, UK\\
$^{2}$Kavli Institute for Cosmology, University of Cambridge, Madingley Road, Cambridge CB3 0HA, UK\\
$^{3}$University of California Observatories - Lick Observatory, University of California Santa Cruz, 1156
High St., Santa Cruz, CA 95064, USA\\
$^{4}$McDonald Observatory, The University of Texas at Austin, 2515 Speedway Stop C1402, Austin, TX 78712, 	USA\\
$^{5}$Departamento de Astronomia, Av. Bento Goncalves 9500, Agronomia, Porto Alegre, RS, Brazil\\
$^{6}$Laborat\'orio Interinstitucional de e-Astronomia, Rua General Jos\'e Cristino, 77 Vasco da Gama,
	Rio de Janeiro, Brazil, 20921-400\\
$^{7}$Departamento de Fisica, Universidade Federal de Santa Maria, CEP 97105-900  Santa Maria, RS - Brazil\\
}

\begin{document}

\date{Accepted . Received}

\pagerange{\pageref{firstpage}--\pageref{lastpage}} \pubyear{2002}

\maketitle

\label{firstpage}

\begin{abstract}
Several models have predicted that stars could form inside galactic outflows and that this would be a new
major mode of galaxy evolution. Observations of galactic outflows have revealed that they host large
amounts of dense and clumpy molecular gas, which provide conditions suitable for star formation.
We have investigated the properties of the
outflows in a large sample of galaxies by exploiting the integral field spectroscopic data of the large MaNGA-SDSS4 galaxy survey.
We find evidence for prominent star formation ocurring inside at least 30\% of the galactic outflows in our sample,  while
signs of star formation are seen in up to half of the outflows.
We also show that even if star formation is prominent inside many other galactic outflows, this may have not been revealed as the
diagnostics are easily dominated by the presence of even faint AGN and shocks. If
very massive outflows typical of distant galaxies and quasars follow the same scaling relations observed locally, then the
star formation inside high-z outflows can be up to several 100 $\rm M_{\odot}~yr^{-1}$ and could contribute
substantially to the early formation of the spheroidal component of galaxies. Star formation
in outflows can also potentially contribute
to establishing the scaling relations between black holes and their host spheroids. Moreover, supernovae exploding on large orbits
can chemically enrich in-situ and heat the circumgalactic and intergalactic medium. Finally, young stars ejected on large  orbits may also contribute to the
reionization of the Universe.
\end{abstract}


\section{Introduction}
\label{sec:intro}

Fast outflows, extending on kiloparsec scales, are commonly found in galaxies hosting enhanced star formation ("starburst") and in
galaxies hosting accreting supermassive black holes (quasars or, more generally, Active Galactic Nuclei, AGN). 
Galactic outflows are often invoked as a negative feedback process in galaxies; indeed, by removing
and heating gas, outflows can potentially suppress star formation in their host galaxies
\citep[e.g.][]{Granato2004,King2010,Fabian2012,King2015}. However, some models have also proposed that outflows and jets
can also have a positive feedback effect; indeed, compressing gas in the interstellar and circumgalactic
galactic outflows can foster the fragmentation and gravitational collapse of clouds, hence boost
star formation \citep{Rees1989,Nayakshin2012,Bieri2016a,Zubovas2013}. However, more recently, models have proposed
an even more fascinating scenario, according to which the outflowing gas should cool, fragment and,
therefore, form stars {\it inside} the outflow
\citep{Ishibashi2012,Zubovas2013a,Silk2013,Zubovas2014,Ishibashi2013,Gaibler2012,Dugan2014,Ishibashi2014,Zubovas2013b,El-Badry2016,Wang2018,Mukherjee2018}.
This new star formation mode is drastically different from star formation in galactic discs and even from
star formation resulting from the ``standard'' positive feedback scenario discussed above.
Indeed, stars formed inside galactic outflows would have high
velocities on nearly radial orbits. Depending on the radius and velocity at the time of formation, stars born inside outflows may
escape the galaxy and/or the halo, or become gravitationally bound \citep{Zubovas2013a}. According to models
this star formation mode can reach a few/several 100 $\rm M_{\odot}~yr^{-1}$ \citep{Ishibashi2012,Silk2013}. The implications of this new
star formation mode are potentially far reaching, ranging from the morphological/dyanmical evoutionary properties of galaxies
to the enrichment of the circumgalactic and intergalactic medium, as will be discussed more in detail in
Sect.\ref{sec:implications}.

Observations have recently shown that fast galactic outflows contain a large amount of cold molecular
gas \citep[e.g.][]{Feruglio2010,Sturm2011,Cicone2014,Garcia-Burillo2015,Combes2014a,Sakamoto2014,Fluetsch2018}, including an
unusually high fraction of dense molecular gas \citep{Aalto2012,Aalto2015b,Walter2017,Privon2017}, which is also found to be highly
clumpy \citep{Pereira-Santaella2016,Finn2014,Borguet2012a}. These observational results
provide strong support to the theoretical scenario in which star formation occur in galactic outflows, since these possess the physical
conditions (cold, dense and clumpy gas) suitable for star formation. 

Several clear examples of star formation triggered in shocked and compressed gas clouds have been unambiguously seen in our Galaxy
\citep{Zavagno2010a,Zavagno2010b,Lim2018,Baug2018,Dwarkadas2017,Duronea2017,Figueira2017,Deharveng2015,Ladeyschikov2015,Dewangan2013,Dewangan2012,
Thompson2012,Brand2011}. On larger scales, evidence for star formation triggered by the interaction of jets or outflows
with the interstellar medium or circumgalactic medium of galaxies has also been detected, although the inferred star formation
rate is generally low ($\rm<1M_{\odot}~yr^{-1}$) \citep{Croft2006,Crockett2012,Santoro2016,Cresci2015a,Molnar2017,Lacy2017,Salome2015,Elbaz2009}.
At high redshift indications of
star formation triggered by outflows or radio jets has also been found, in some cases potentially reaching a few hundred
$\rm M_{\odot}~yr^{-1}$ \citep{Cresci2015b,Bicknell2000}. However, all these cases are examples of the ``standard'' positive
feedback mode, in which star formation is triggerd as a consequence
of the compression of the
insterstellar/circumgalactic medium resulting from the interaction with the jet/outflows.

Evidence for the new mode of positive feedback, i.e. star formation {\it inside} galactic outflows, has been found only recently
by \cite{Maiolino2017}. They analysed optical/near-IR spectroscopic data of the prominent galactic outflow previously detected
in the galaxy IRAS F23128-5919 \citep{Leslie2014,Bellocchi2013,Arribas2014,Cazzoli2016,Piqueras-Lopez2012} and found
that multiple optical and near-IR diagnostics of the outflowing gas are consistent with {\it in-situ} star formation, i.e.
inside the outflow. However, while this has been a remarkable result, it has not been clear whether this is an isolated, rare
case, or star formation is more common in galactic outflows, but it has been difficult to identify.

Indeed, outflows are often driven by AGNs, which (as we shall discuss in Sect.\ref{sec:agn_dominance}) tend to dominate the diagnostics,
especially in integrated or nuclear spectra,
and the presence of shocks may also hinder the detection of star formation.

We have investigated the occurrence of this phenomenon in a systemic and unbiased way by exploiting the integral field spectra of
the 2,800 galaxies of the MaNGA-SDSS4 DR2 survey. These offer high quality, spatially resolved spectra spanning
a broad wavelength range, hence enabling us to probe multiple diagnostics. We identify a subsample of 37 galaxies which
show evidence for clear outflows and which can be traced thorugh all primary nebular lines required for identifying
the sources of gas ionization/excitation through diagnostic diagrams. We have therefore exploited these data to map
the diagnostics across each outflow, hence revealing the presence of star formation inside outflows. As we will discuss
in detail, we have found that about one third of outflows are characterized by prominent star formation inside them,
and about half of the outflows show evidence for at least some star formation, hence revealing that star formation
is common to most galactic outflows, with major implications especially for high redshift galaxies, where outflows
are much more prominent.

In this paper we assume the following cosmological parameters:
$\rm H_0 = 70~km~s^{-1}~Mpc^{-1}$, $\rm \Omega_\Lambda = 0.7$
and $\rm \Omega _m = 0.3$.

\section{Data analysis}
\label{sec:analysis}

\subsection{The MaNGA survey}
\label{sec:manga}

The observations for the SDSS-IV MaNGA survey 
\citep{Bundy2015,Blanton2017}, are taken on the Sloan 2.5m telescope \citep{Gunn2006},
using hexagonal IFU fibre bundles, ranging from 19 to
127 fibres in size, with 17 galaxies observed simultaneously.
These IFUs feed the light into the dual-channel BOSS spectrographs
\citep{Drory2015}.
The size of the fibre bundle is chosen to ensure coverage out to a minimum of 1.5 Re for the MaNGA primary sample, and 2.5
Re for the secondary sample. The observing strategy chosen utilises a 3-point dithering pattern, with a total exposure time per
galaxy of around 3 hours, aiming to reach the MaNGA requirement of signal-to-noise of 5 per spectral pixel in the r-band continuum
at a surface brightness of 23 AB arcsec$^{-2}$. Each set of exposures is also required to provide a median seeing of 2.0 arcsec or
below, providing a PSF of less than 3 arcsec in the reconstructed images \citep{Law2015,Law2016,Yan2016a,Yan2016b}.

The wavelength coverage in the MaNGA cubes ranges from 3600\AA\  to 10300\AA\  at R$\sim$2000, providing simultaneous access to the nebular
lines [OII]3727,3730, H$\beta$, [OIII]4959,5007, [OI]6300,6366, [NII]6548,6584, H$\alpha$
and [SII]6717,6731. These are fundamental nebular lines
which can be used to investigate the excitation and physical properties of the ionized gas. In particular,
the so-called ``BPT'' diagnostic diagrams \citep{Baldwin1981,Kewley2006} compare the ratios of different nebular emission lines
(specifically: [OIII]$\lambda$5007/H$\beta$ versus [NII]$\lambda$6584/H$\alpha$, [SII]$\lambda$6717+6731/H$\alpha$ and [OI]$\lambda$6300/H$\alpha$)
to enable a classification in terms of sources responsible for ionizing and exciting the gas, specifically
differentiating between excitation from young hot stars in star forming regions and other forms of excitation such as
AGNs, evolved stellar populations and shocks.

\begin{figure*}
\centerline{\includegraphics[width=10.5truecm,angle=0]{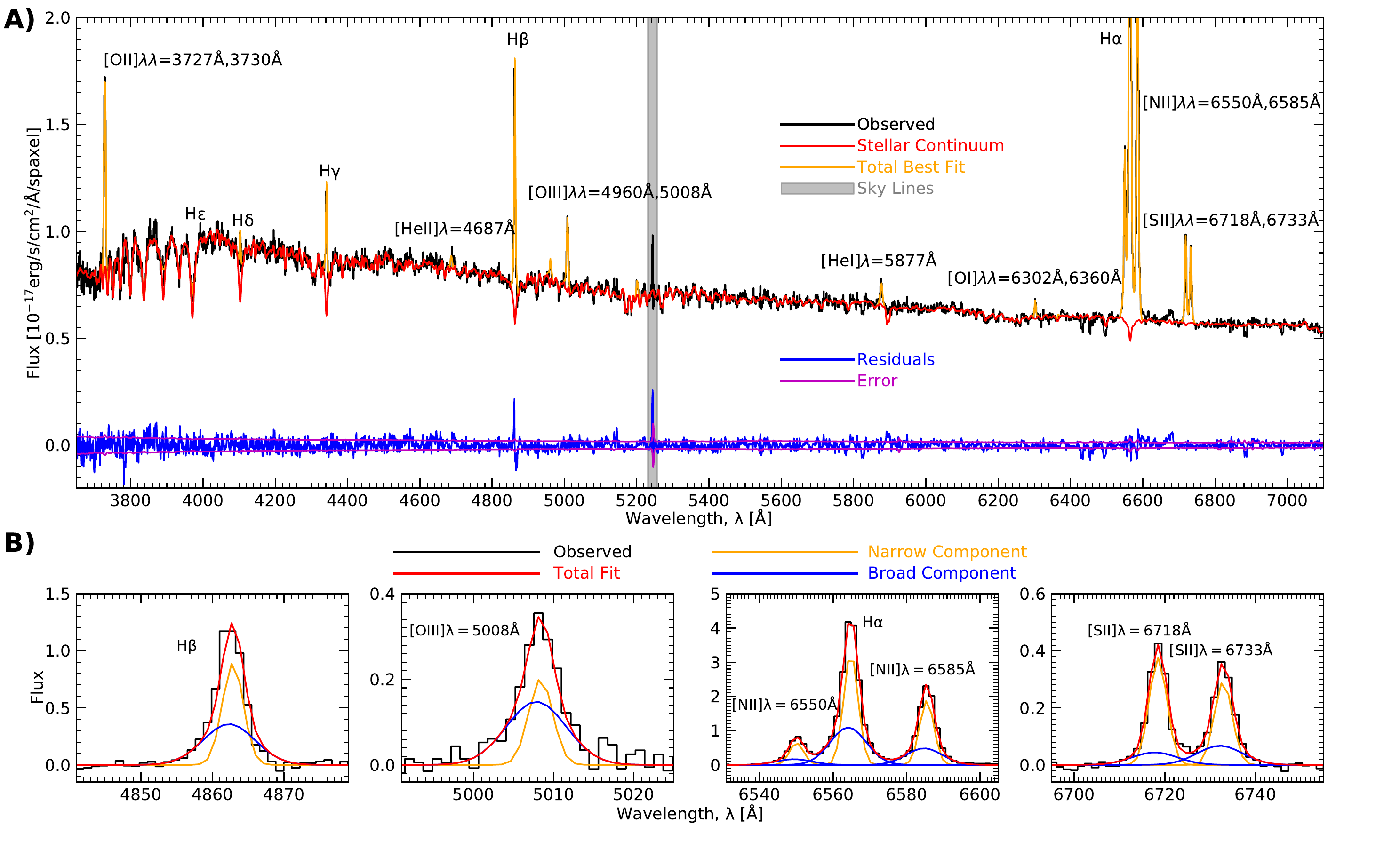}}
\caption{Example of simultaneous stellar continuum and emission line fit to the spectrum extracted
from the central region of a representative galaxy in our sample. A: Full spectrum decomposition showing the
simultaneous stellar continuum (red) and emission line (orange) fit to spectrum. Shown
below the fit are the residuals (blue) and the error spectrum (magenta) for the galaxy. The OII 5577 sky line has been masked, shown
here by the grey shaded area. B: Subsections of continuum-subtracted spectra, showing some of the relevant emission lines used in
the analyses. The decomposition of the narrow (orange) and broad (blue) components can be seen, alongside the total fit of the two
components (red).}
\label{fig:spectrum}
\end{figure*}

\subsection{Spectral fitting, identification and mapping of outflows}
\label{sec:spectral_fitting}

We have considered all 
IFU spectra of the 2,800 galaxies in the MaNGA DR2. Since outflows (both AGN-driven and SF-driven)
generally are more prominent in the central region
of galaxies) we have first inspected the spectrum of the central region of the galaxy,
i.e. extracted from the central 2.5" (which corresponds to the typical PSF of the Sloan 2.5m Telescope),
with the goal of identifying candidate galaxies showing indication of outflows. One aside to the decision for this scale is that, due to the 
MaNGA galaxies spanning a range of redshifts, this area from which we extract our "central" spectrum will correspond to a different 
physical scale on each galaxy.

The first step was to fit the stellar continuum in this central region. For this purpose, we used the full MILES empirical stellar
library, consisting of 985 stellar templates. Unfortunately, this library only covers the wavelength range 3525-7500\AA, hence not
fully exploiting the full coverage of the MaNGA data, which extends up to 10,000\AA. However, it covers the continuum beneath all
nebular lines of interest, and has a spectral resolution high enough to match the MaNGA spectra. The fitting of the continuum was
done by using an adapted version of the PPXF routine \citep{Cappellari2017}.

We begin the analysis of the central spectrum by fitting the stellar continuum and kinematics, whilst also masking all nebular
emission lines within $\pm$1000~km~s$^{-1}$ from each line's redshifted central wavelength. The continuum is fit using the PPXF routine,
assuming four Lines Of Sight Velocity Distribution (LOSVD) components (v, $\sigma$, h3 and h4), and allowing a 6th order additive
Legendre polynomial to correct the continuum shape for small residual background. The OII 5577 sky line is also masked, as although good sky subtraction has
been performed on the MaNGA data, some residuals can still exist for this line. 

Having fit the stellar continuum, the residual spectrum from the fit was next tested for multiple emission line components. To
investigate the presence of multiple components in the emission lines we began by testing the primary nebular lines mentioned
above. 
We first simultaneously fit all nebular lines with a single Gaussian component, imposing the same central velocity and width for
all nebular lines. The intensities of the Gaussian components of the individual nebular lines were left free to vary, with the
exception of the [NII], [OIII] and [OI] doublets, whose line ratios were tied by their Einstein coefficients. 

Note that when tying the velocity dispersions, one has to take into account that the spectral resolution changes slightly across
the MaNGA spectrum. However, the latter is very well characterised in each spectral region and provided together with each
spectrum. Therefore, at each point we can link the intrinsic velocity dispersion/width (which is to be tied for all lines) to the observed velocity
width through the relation
$$ \sigma _{line} = \sqrt{\sigma ^2_{SRES}+\sigma ^2_{int}}$$
where $ \sigma _{line}$ is the dispersion of the final Gaussian fit, $\sigma _{SRES}$
is the spectral resolution at the observed wavelength of the line and $\sigma _{int}$ is the tied dispersion between all lines.

Following this, we then investigated the presence of a second, broader component of the nebular lines. Assuming the emission lines
could be well modelled by Gaussian templates, we allowed up to two Gaussians to be fit to each emission line, fitting only the first
two orders of the LOSVD (i.e. velocity and dispersion).

In type-1 AGNs the very broad emission lines from the Broad Line Region (BLR) make it difficult to characterize the outflow,
therefore we have decided to discard this class of galaxies.
Type-1 AGNs were identified as galaxies with nuclear very broad hydrogen Balmer lines  (typically with 
FWHM ranging from 1,000km/s to 10,000km/s) without an associated broad component of the forbidden transitions (in particular [OIII] and
[SII]), which indicate that the broad Balmer component is indeed associated with the BLR and not outflow.


The putative additional (broader) Gaussian component was also required to have the same velocity and width for all nebular lines,
with only the intensity of the broad component of the individual nebular lines left free to vary, though with the same constraints
on the line doublets as discussed above. For our pipeline to accept a second component, we require a number of conditions to be met.
The most basic condition is that the multiple components fit improved upon the reduced chi-squared of the single component
fit by a minimum of 10\%. This is always a very conservative requirement, as with so many spectral points this always translates into
a statistical significance of the additional component higher than 99\%. In principle, one could relax this requirement, but, in this
paper, we have preferred to be conservative in the identification of the broad components. 

We also place two restrictions upon the broad components for both the H$\alpha$ and [OIII]5007 lines. Firstly, we require that the 
signal-to-noise ratio (defined by the peak flux of the fitted Gaussian divided by the median error within the FWHM of the line)
is greater than 3. Secondly, we require that the Equivalent Width of the broad component (relative to the total continuum) must also be greater than 3. These
requirements are put in place to ensure that the broad component is a distinct feature, separate from the narrow component and
not simply an artefact of poor stellar continuum fittings. In a later discussion within this paper, we mention how we cannot disentangle the
stellar continuum of the constituent stars in the galaxy from those being formed within outflows, preventing us from separating the true
Equivalent Width associated with the outflowing (broad) components.
However, in the context of this section we are now mostly focused on the operational identification of the broad component and making
sure that it is not resulting, as already discussed above,
from an inappropriate subtraction of the stellar continuum, hence we consider the equivalent width of the
broad component relative to the total stellar continuum.
In many of the
cases, the restriction on the signal-to-noise is strict enough that spurious sources are removed without consideration of the Equivalent
Width regardless. In practice, for this specific paper, these requirements are overridden by the subsequent requirements of detecting
also the broad component of the fainter lines required for the BPT classifications. Any galaxy that exhibits a broad component which
meets both of these requirements is considered to be an outflow candidate for further consideration.

The broad component of [OI]6300 is often very weak and often undetected. However, generally, when non-detected the upper limit on [OI]6300 is meaningful, in the sense that it locates the broad component in the HII-like (star forming) region of the [OI]-BPT diagram.

Having positively identified multiple components in a spectrum, we next performed a check to ensure the emission line masks provided
in the first steps are sufficiently masking the emission lines. We define a line a being sufficiently masked if the mask covers out
to 3$\sigma$
from each line centre (using the updated values of line centres and widths). Insufficient masks can lead to poor recovery of
the stellar kinematics. Thus, if any of the lines are broader than their initial masks, then the masks are moved and increased in
size to sufficiently mask the line, and the stellar kinematics are refit as before, and emission lines tested again. This step is
repeated iteratively until the reduced chi squared of the continuum fit does not improve upon the previous fit's reduced chi
squared. At this point, the stellar kinematics are assumed good, and are saved and held fixed in all future steps. 

In the next stage of the fitting process, the stellar continuum is fit again, with fixed kinematics, whilst simultaneously fitting
the emission lines, whose kinematics were free to vary. This method was chosen as it has been previously shown that line fluxes can
be underestimated by as much as 10\% when fit post continuum subtraction \citep{Sarzi2006}.
With some of our lines being relatively low flux
(especially for H$\beta$), we indeed found that in cases where the fit was performed post subtraction, the continuum fit itself was
sometimes modified by PPXF to account for template mismatch through the additive polynomials used to correct for continuum shape. In
these situations, the simultaneous fitting of the continuum and the emission lines alleviated the problem.

\begin{figure*}
\centerline{\includegraphics[width=13truecm,angle=0]{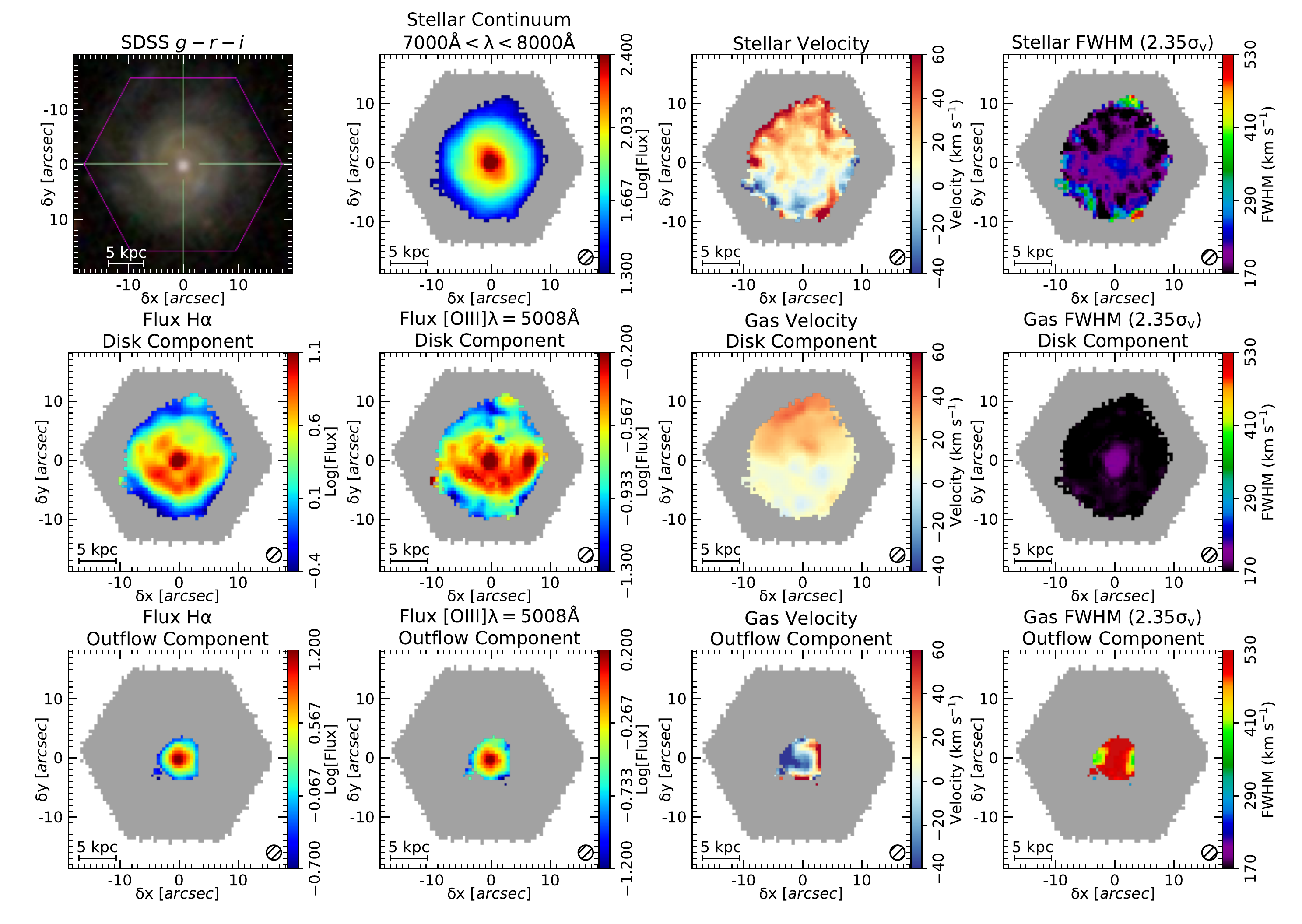}}
\caption{Example multi-component mapping. For the same representative galaxy of
Fig.\ref{fig:spectrum} the panels show flux, velocity and velocity dispersion
maps for the stellar population, for the narrow component of the nebular lines,
tracing the galactic disc, and for the broad component of the nebular lines,
tracing the outflow. The grey regions indicate spaxels within the
Manga fibre bundle footprint but masked because either the stellar continuum or
emission line component is not detected.}
\label{fig:map}
\end{figure*}

Figure
\ref{fig:spectrum} shows the
spectral fitting of the central region of a representative galaxy resulting from the steps described above.

For those galaxies showing evidence for an outflow in the central spectrum, we then extend the map
of the associated broad components across the entire galaxy.

We begin by estimating the continuum SNR of each spaxel. This was determined by taking the medians of the flux and inverse variance
spectra, provided in the MaNGA cubes, within the SDSS R-band (5560 \AA $< \lambda <$6942 \AA). All spaxels with SNR $<$ 5 were removed at this point, as the spaxels were to be binned to higher signal noise, and these spaxels would lead to large bins with noisy data.

Taking the remaining spaxels, all of the spectra were first collapsed into a single integrated spectrum. This was fit using similar steps as discussed above, and the non-zero weighted templates extracted to create individual template libraries for each galaxy, typically reducing the full MILES
library from 985 to 25-40 templates that best described the galaxy properties. This was done to speed up the fitting process, typically by an order
of magnitude.

Next, the spaxels were Voronoi binned to increase the SNR of the continuum for the binned spectra to 25. Through the use of Monte-Carlo simulations,
it was determined that a minimum SNR of 25 should be used for a sufficiently accurate recovery of the stellar kinematics \citep{Sarzi2006}.
The high SNR binned spectrum is used to restrict the number of templates to be used for the stellar fitting
within the bin and to provide the initial parameters for fitting the stellar kinematics and for the nebular lines. Then the latter information is used
to fit each individual spaxel within the Voronoi bin, by following the same steps discussed above for the central spectrum. 

Figure \ref{fig:map}A shows the kinematics of the narrow (disc) and broad (outflow) components, along with the stellar kinematics,
for the same representative galaxy analysed in Figure \ref{fig:spectrum}. The same figure shows the flux distribution of both the narrow and
broad components of both the H$\alpha$ and [OIII]5007 lines. The same flux maps have been obtained also for the other lines required for the
BPT diagram.

This analysis could be performed for a total of 37 MaNGA galaxies. This does not mean that other galaxies in the MaNGA data
do not show outflows, but that only for these 37 galaxies were all of the lines required for BPT analysis prominent, and that the 
outflowing gas could be kinematically separated from the constituent gas of the galaxy disk, allowing for a clean and detailed analysis
of the outflow using BPT diagnostics.

The list of the 37 galaxies with outflows that can be characterized through the BPT diagnostics is given in Table~\ref{tab:z_mass}, along
with their stellar masses from the MPA-JHU catalogue \citep{Brinchmann2004}.

The fitting results and maps (including the BPT maps for the narrow components)
for the full set of 37 MaNGA galaxies for which this analysis has been possible are available
in electronic form in the accompanying supplementary online material.

Finally we mention that the fluxes of the nebular emission lines, both narrow and broad, were corrected for dust reddening by assuming
the default \cite{Calzetti2000} attenuation curve and using the H$\alpha$/H$\beta$ ratio to infer reddening. 

\begin{figure*}
\centerline{\includegraphics[width=13truecm,angle=0]{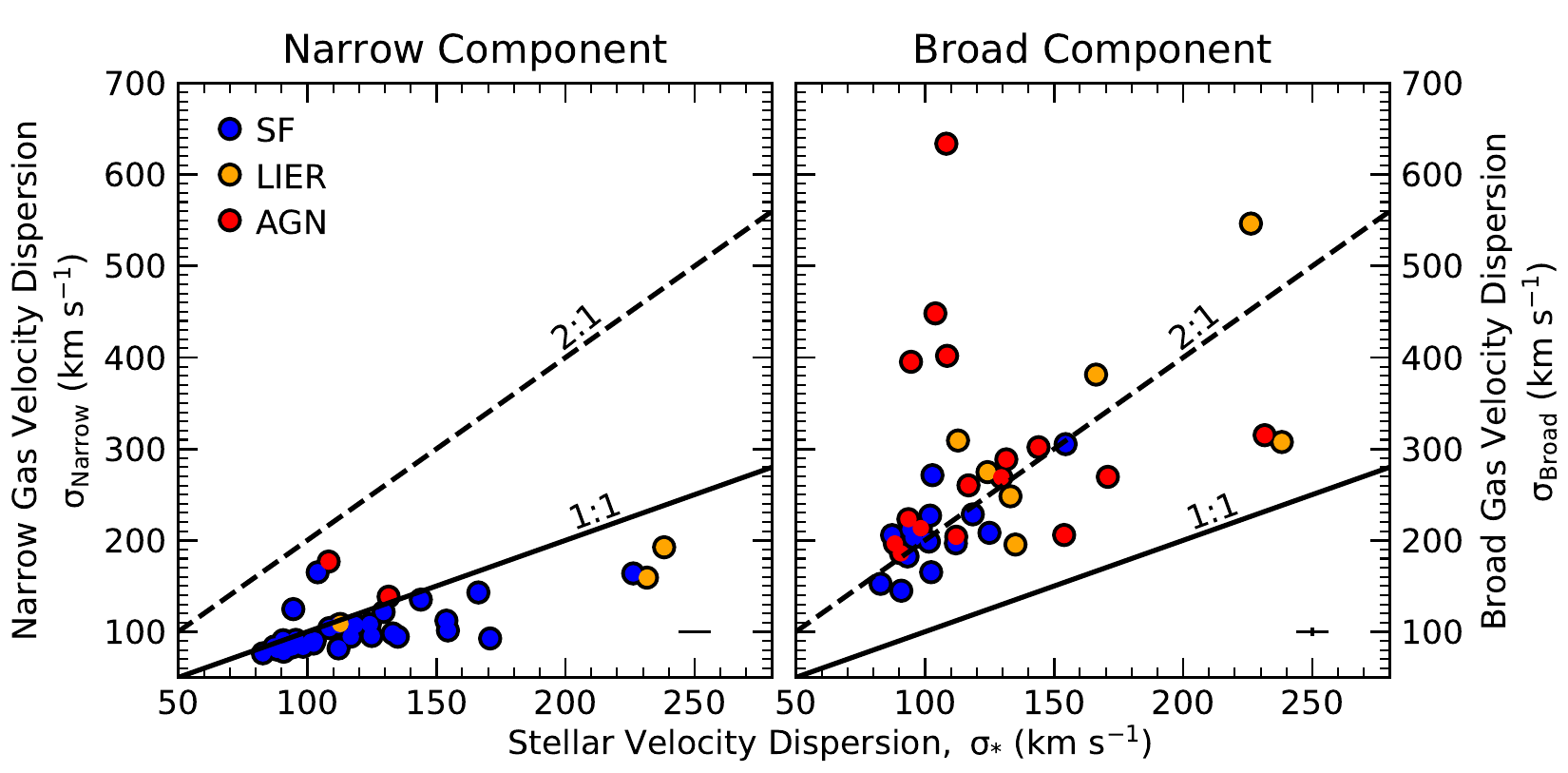}}
\caption{Left: Velocity dispersion of the narrow
component of the nebular lines in the central region compared with the stellar velocity dispersion. As expected, the narrow component has a velocity
dispersion similar to the stellar velocity dispersion, or even lower, owing to the fact that the gas disc is generally dynamically colder than
stars. Right: Velocity dispersion of the broad component of the nebular lines in the central region compared with the stellar velocity dispersion in the
same region. The former is much larger than the latter, indicating that the broad component cannot result from beam smearing effects of the central
rotation curve and that it must be associated with non-virial motions, i.e. outflows.}
\label{fig:sigma_gas_stars}
\end{figure*}

\subsection{Investigating the effects of beam smearing}
\label{sec:beam_smearing}

A potential problem of the modest angular resolution of the MaNGA data is that the putative broad component in the innermost, central region may
simply be an artefact of beam smearing of the rotation in the central region. Many authors explore this issue by modelling the rotation curve and
then simulating the effect of the beam smearing; however, such a test is dependent on the model adopted to describe the velocity field in the central
region; therefore, to avoid being model-dependent we have not used this method.

Instead, we have simply compared the velocity dispersion of the broad component in the central region with the central stellar velocity dispersion,
which is affected by the same beam smearing. A broadening resulting from beam smearing of the stellar velocity field should result in the stellar
velocity dispersion being as high as the broad component of the gas. Figure \ref{fig:sigma_gas_stars}
shows comparisons of the central velocity dispersion of both the
narrow (left) and broad (right) gas components with the stellar velocity dispersion.
The narrow components tend to show a similar dispersion as the stars (the
solid line represents the 1:1 ratio), and in some cases a smaller dispersion, as the gaseous disk is typically dynamically cooler than the stars in
the bulge \citep[e.g.][]{Bertola1995,Corsini1999,Young2008,Martinsson2013}. Conversely, the broad components are always found to have a dispersion much larger than the stellar component (the dashed line represents
the 2:1 ratio), indicating that the broadening cannot originate from beam smearing. This further indicates that the broad component is tracing gas
not in virial equilibrium with the gravitational field, which is traced by the old stellar population, hence implying that the broad component is truly tracing
outflowing material.  A further indication that the broad component is not resulting from effects of beam smearing is the fact that such component
is not distributed along the minor axis of rotation (which instead is a typical signature of broadening by beam smearing of the rotation field)
while it is well extended also fully resolved along the major axis direction.
The fact that
the outflow is resolved is further confirmed by the fact that the broad components have clear kinematic structures both in the velocity map and in
the dispersion map.

\begin{figure*}
\centerline{\includegraphics[width=14truecm,angle=0]{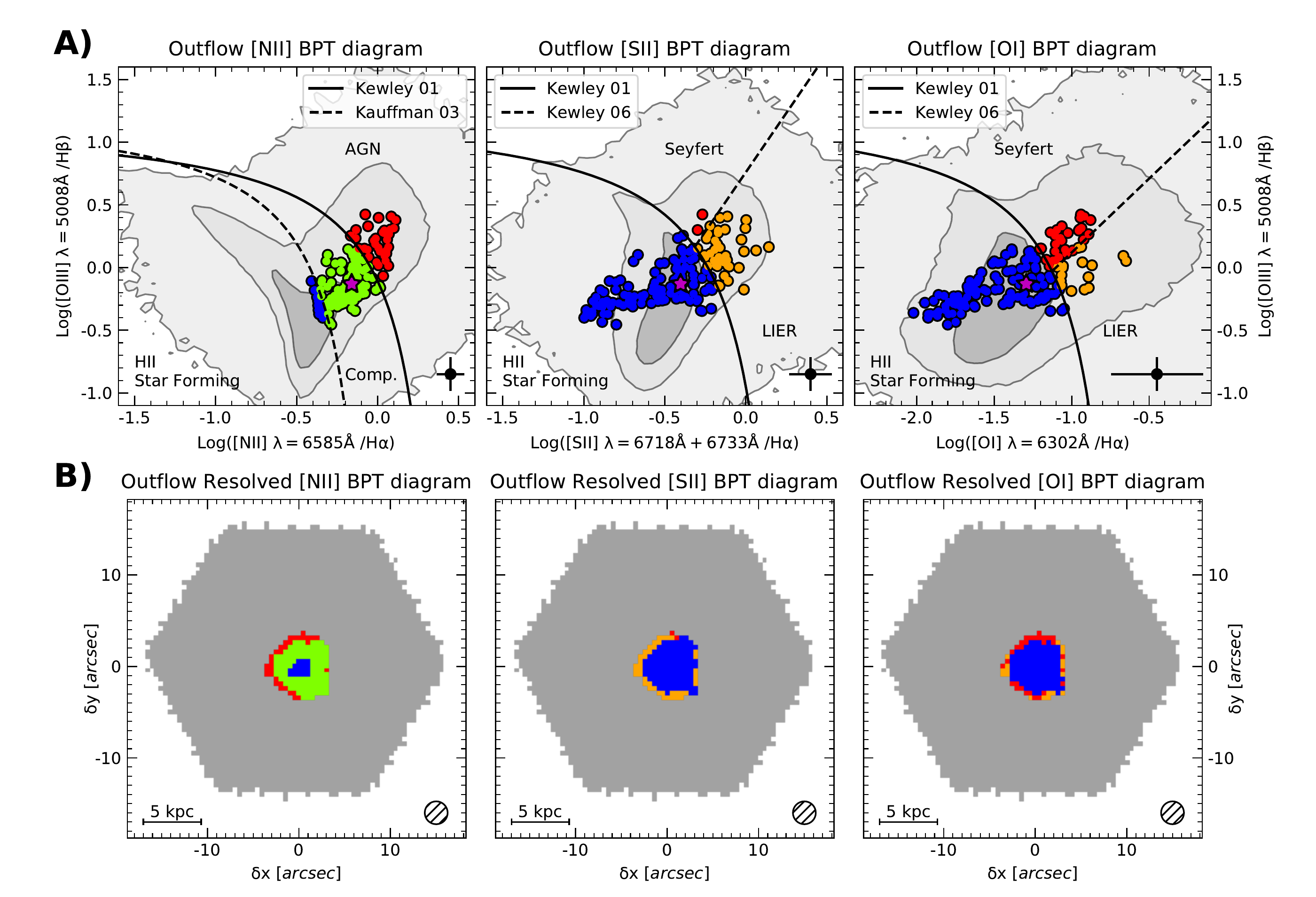}}
\caption{A: Distribution of the broad (outflow) component of the nebular lines on the BPT diagrams for the same representative galaxy of
Fig.\ref{fig:spectrum} and Fig.\ref{fig:map}. A significant fraction of this galactic outflow is in the BPT regions typically populated by star
forming regions and star forming galaxies. The magenta star shows the median location of the BPT points in the outflow. The background
grey-scale contours indicate the distribution of several hundred thousand galaxies in the SDSS survey. B: Spatially resolved BPT classification
of the outflowing gas, using the same colour coding as in (A). The central region of the outflow is mostly star forming, while towards the outer parts of the outflow excitation by shocks likely dominates.  Maps for the other 37 galaxies, including BPT maps for the narrow components, can be found in the
supplementary material.}
\label{fig:bpt_map}
\end{figure*}

\begin{figure*}
\centerline{\includegraphics[width=18truecm,angle=0]{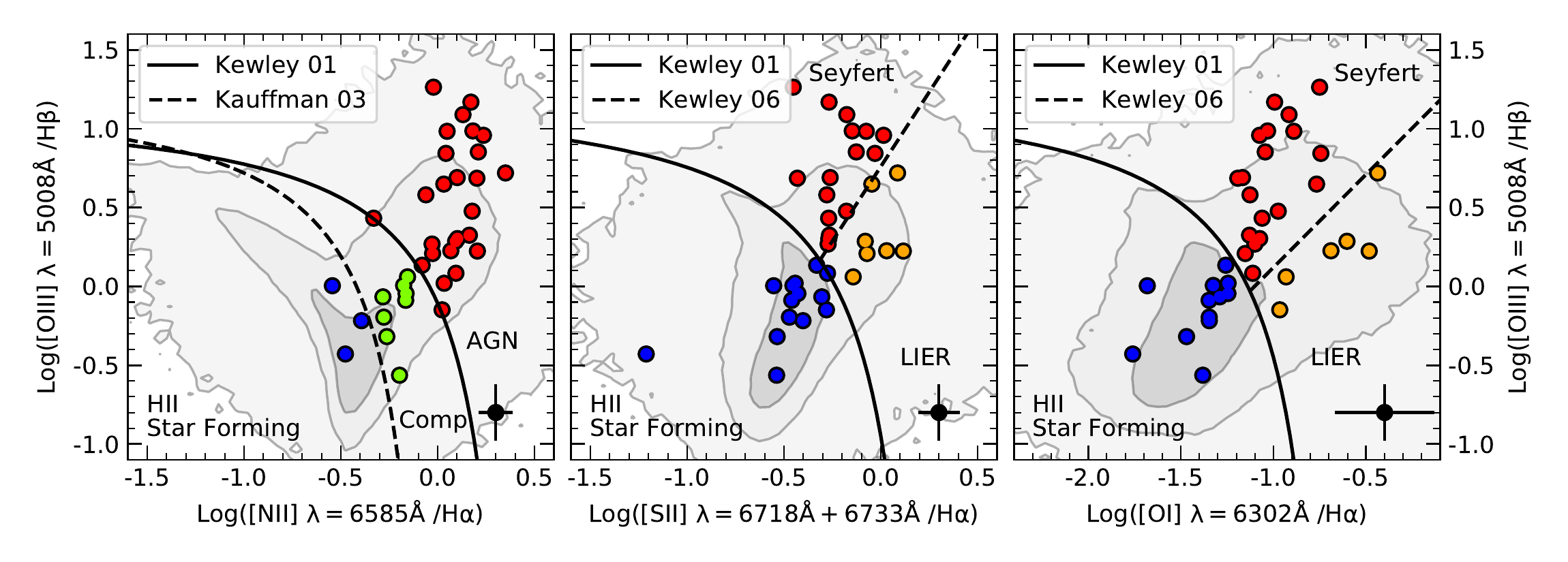}}
\caption{Median BPT classification of the outflows for the galaxies in our sample. According to their median BPT classification in the [SII] and [OI]
diagrams (central and right panes), about 30\% of the galactic outflows in our sample are ``star forming''.
The same applies to the [NII] classification if one includes galaxies classified as ``composite'' (however, one should take into account that the [NII] classification is more ambiguous, as discussed in the text).}
\label{fig:bpt_total}
\end{figure*}

\begin{figure*}
\centerline{\includegraphics[width=17truecm,angle=0]{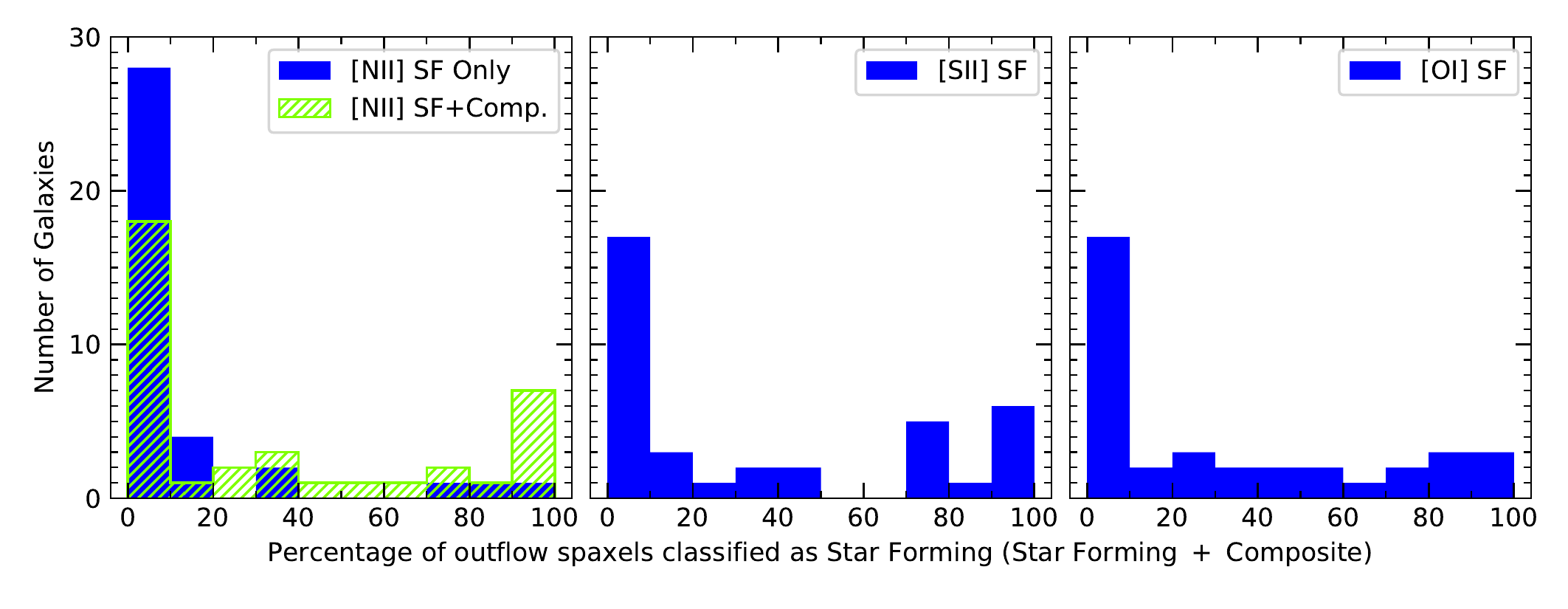}}
\caption{Distribution of the outflows as a function of their star forming fraction. Each panel shows the distribution of outflows which have a given
fraction of spaxels classified as star forming according to each of the three BPT diagrams. In the case of the [NII]-BPT the combined case of "star
forming" and "composite" is also shown. According to the [SII] and [OI] classifications about half of the outflows have at least 10\% of the spaxels
that are star forming. The same result applies to the [NII] classification if one includes the composite cases.}
\label{fig:bpt_frac}
\end{figure*}

\section{Fraction of outflows hosting star formation} 
\label{sec:sf_in_outflow}

\subsection{Spatially resolved BPT-diagnostics}
\label{sec:bpt_spatial}

We have explored the properties of the outflow by investigating the spatially resolved BPT diagrams of the outflow in each galaxy.
We recall that these diagrams enable the classification of galactic regions in terms of gas excitation/ionization mechanisms, specifically:
star formation (i.e. HII region excited by young, massive hot stars), excitation by supermassive accreting black holes (i.e. Seyfert or Quasar
nuclei), and Low Ionization Nuclear Emission Line Regions (LINERs). The latter is a class of object comprising a broad range
of excitation mechanisms: initially it was proposed that they are associated with excitation
by weak, radiatively ineffcient accreting black holes \citep[e.g.][]{Ho1993}, however it has been realized that in most local galaxies
the ``LINER'' emission is generally extended on
kpc-scales across a large fraction of passive and green valley
galaxies (hence renaming this class as ``LIER'', i.e. dropping
the ``N'' which stands for ``Nuclear'' in the original acronym)
and correlates with the old stellar population, and this can
be explained in terms of excitation by the hard radiation
field produced by evolved post-AGB stars \citep[e.g.][]{Sarzi2010,Belfiore2016a}. However, in some active galaxies and outflows
the LI(N)ER emission can also be associated with shocks excitation \citep[e.g.][]{Heckman1987,Monreal-Ibero2010}. A more detailed discussion
on the origin of the LI(N)ER emission goes beyond the scope of this paper.

Figure \ref{fig:bpt_map}A shows the line of demarcation between
the different excitation mechanisms
identified by \cite{Kauffmann2003a}, \cite{Kewley2001} and \cite{Kewley2006}, in the three BPT diagrams.
The gray-shaded contours show the distribution of several hundred thousands galaxies in the SDSS (single fibre) DR7 survey \citep[the distribution
of spatially resolved MaNGA spaxels for the full survey is broadly consistent with the single fibre DR7 data, however we prefer
to use the latter as the MaNGA spaxels underpolulate the AGN locus, ][]{Belfiore2016a}.
On the [NII]-BPT diagram the Seyfert/quasars and LI(N)ER are not clearly divided and are generally classified together as ``AGN''.
On the same diagram \cite{Kauffmann2003a} identified a demarcation curve between HII/Star-forming regions and AGNs (dashed line) empirically
based on the distribution of galaxies in the SDSS survey, while \citep{Kewley2001} provided a curve (solid line) tracing the maximum line ratios
associated with star forming regions according to photoionization models; galaxies/regions in the area between these two curves are often
refereed to as ``composite'' systems, in which both SF and AGN probably coexist.
However, we shall note that objects in this region are still fully
compatible with the maximum line ratios expected from star forming galaxies,
hence they can still be fully star forming.

We note that other diagnostic diagrams involving the equivalent width of H$\alpha$
\citep{Cid-Fernandes2011}
cannot be used in this case, as we cannot discriminate the continuum associated with the broad and narrow components. Therefore,
in this paper we only expoit the BPT diagrams to identify the excitation mechanism in galactic regions and specifically in the outflows.

As an example of BPT diagnostics spatially resolved in outflows,
Figure \ref{fig:bpt_map}A shows the BPT classification of the
broad (outflow) component of each individual spaxel of the same representative galaxy shown in representative galaxy Figure \ref{fig:map}.
The representative error bars shown in the bottom right show the
median errors for each of the ratios (where the errors are determined from the errors returned on the fit to the velocities, velocity dispersions
and fluxes for each line.
Figure \ref{fig:bpt_map}B shows how the BPT classification is distributed within the outflow map.
For this galaxy, both the [SII]-BPT and [OI]-BPT diagrams consistently indicate that
the central region of the outflow is star formation-dominated, while shocks likely dominate in the outer regions of the outflow.
The [NII]-BPT diagram also provides a consistent picture if one includes
``composite'' spaxels (green) as hosting star formation, since, as
discussed above, objects in this region of the BPT diagram can still
potentially be fully star forming.
One should also take into account that the [NII]-BPT diagram is considered as the least reliable of such diagrams because of
its strong dependence on the nitrogen abundance \citep{Masters2016} and on the ionization parameter \citep{Strom2017b}. 

To have a useful quantities of comparison for the outflow in the different galaxies in our sample,
we condense the information on each outflow by taking
the median location on the BPT diagram of such outflow components for each galactic outflow. In the case of the galaxy shown
in Figure \ref{fig:bpt_map}A the median location in each BPT diagram is shown with a magenta star symbol.
We note that this is a considerable improvement, enabled by spatially resolved spectroscopy, relative to spatially integrated spectra,
which are generally light-weighted and therefore dominated by the brightest regions (e.g. a central AGN-dominated region).

Figure \ref{fig:bpt_total} shows the resulting median location of the outflows on the BPT diagrams for all systems in our sample (Table \ref{tab:sfr_outf} gives
the average BPT classification for each indivdual object). The [SII]-BPT and
[OI]-BPT diagrams indicate that in most outflows the ionized gas appears to be excited by AGN, however, about 30\% of the outflows host
gas whose ionization is dominated by young stars. For the [NII]-BPT a smaller fraction of outflows appears dominated by star formation
excitation. However, if one also includes outflows whose [NII]-BPT classification is in the ``composite'' region (considered 
as a transition region in which star formation and AGN excitation coexists), then the fraction of outflows showing indication
of star formation increases to about 30\% also in the [NII]-BPT diagram. One should also take into account the concerns about
the [NII]-BPT diagram discussed above.

Even for those galactic outflows whose median BPT diagnostics are associated with AGN or LIER-like
ionization, there are generally still regions of the outflow that are associated with star formation. Therefore, to better quantify the
occurrence of star formation in galactic outflows, we have estimated  for each outflow the fraction of spaxels that can be classified as "star
forming" according to each of the BPT diagnostics. Figure \ref{fig:bpt_frac} shows the distribution of galaxies as a function of the fraction of the outflow that
is classified as star forming. According to the [SII]-BPT and [OI]-BPT classifications, about half of the galactic outflows show evidence for
some star formation inside the outflow (whose contribution to the overall excitation of the gas ranges from 10\% to 100\%). Even the [NII]-BPT diagram
provides the same fraction of ``star forming outflows'' if one considers also outflowing regions classified as "composite" in this diagram.

\subsection{In-situ versus external photoionization}
\label{sec:u}

One potential concern of our finding could be that, for those outflowing regions where the BPT diagnostics of the outflowing gas are consistent
with excitation from young stars, the latter may not be located inside the outflow but in the galactic disc and the outflowing gas may simply be
illuminated externally by the UV radiation coming from the disc. However, these scenarios can be distinguished by exploiting the ionization
parameter, defined as ratio between ionizing photons flux ($\rm Q_{ion}/4\pi r^2$) and gas electronic density ($\rm n_e$),
and normalized through the speed of light, 
$$\rm U = \frac{Q_{ion}}{4\pi r^2 c~n_e}$$

\begin{figure}
\centerline{\includegraphics[width=9truecm,angle=0]{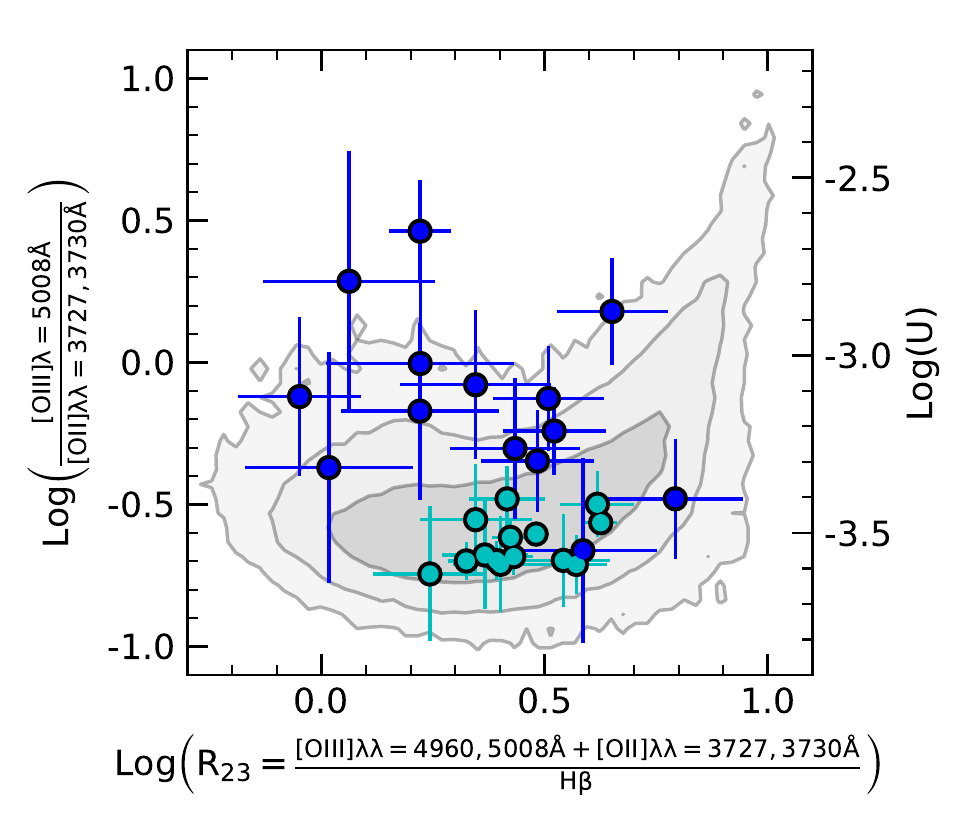}}
\caption{Ionization parameter (right hand Y axis) as traced by the [OIII]/[OII] ratio (left hand Y axis) as a function
of the R$_{23}$ parameter, which is primarily sensitive to the gas metallicity.
Dark blue symbols indicate the median values observed in those galactic outflows dominated by star formation, according to the[SII]-BPT diagnostic diagram, as in Figure
\ref{fig:bpt_total}. The light blue symbols are the average location of the narrow component in regions that are classified as star forming.  The shaded contours
indicate the distribution of star forming galaxies from the SDSS survey (68\%, 95\% and 99.7\% of the population).
Star forming outflows (blue symbols) have similar ionization parameter as normal star forming regions, and possibly even slightly higher.
The error bars shown are the median error of the ratios calculated for all spaxels within the outflow.}
\label{fig:u}
\end{figure}

The gas density in the outflow is similar, or even higher than in the disc, as inferred from the [SII] doublet (Gallagher et al. in prep.).
Therefore, in the scenario of external photoionization one would expect that the ionizing flux should be much lower than in the
case of {\it in-situ} photoionization, therefore resulting into an ionization parameter orders of magnitude lower than observed in standard
star forming regions.

The ionization parameter can be traced through the [OIII]$\lambda$5007/[OII]$\lambda$3727 line ratio \citep{Diaz2000}. This ratio has however
also a secondary dependence on the gas metallicity. The latter can however be monitored through the R$_{23}$ parameter defined as
the ratio $\rm ([OIII]_{4960+5007}+[OII]_{3727})/H\beta$, which is primarily sensitive to the oxygen abundance, and with a secondary dependence
on the ionization parameter. Thefore, diagrams with both these two quantities are used to help disentangling these properties
\citep[e.g.][]{Nagao2006a}.

Figure \ref{fig:u} shows the [OIII]$\lambda$5007/[OII]$\lambda$3727 ratio as a function of the R$_{23}$ parameter. The right-hand axis
translates the [OIII]$\lambda$5007/[OII]$\lambda$3727 ratio into U following the relation provided by \cite{Diaz2000}. The shaded contours
indicate the distribution of star forming galaxies from the SDSS survey (68\%, 95\% and 99.7\% of the population),
illustrating the well-known correlation between U and metallicity \citep[e.g.][]{Nagao2006a}.
Dark blue symbols indicate the median values observed in the galactic outflows of our sample that are dominated by star formation, according to the [SII]-BPT diagnostic
diagram, as in Figure
\ref{fig:bpt_total}, while light blue symbols show the medial location of the narrow component tracing star formation in the galaxy disc.
Figure \ref{fig:u} shows that the
ionization parameter of the gas in the star forming outflows, as inferred from the [OIII]$\lambda$5007/[OII]$\lambda$3727 line ratio,
is undistinguishable from
that of normal star forming regions/galaxies and, if anything, is even slightly higher. This result
excludes that the photoionization in these outflows is due to external UV radiation coming from the star forming disc, and
confiriming that the outflowing gas is excited by {\it in-situ} (i.e. within the outflow) star
formation.


\subsection{Contribution by shocks}
\label{sec:shocks}

One caveat to using both the BPT diagrams and the R$_{23}$ vs O$_{32}$ diagram in identifying in-situ star formation lies in the fact
that, although generally shocks populate the LIER-like region of these diagram \citep[e.g.][]{Allen2008,Diniz2017}, some peculiar shocks (especially low velocity ones) can potentially
produce star-formation-like emission line ratios on the BPT
diagrams. In \cite{Maiolino2017} this form of potential shock excitation was excluded through the use of IR diagnostics, namely the 
[Fe II]$\mathrm{\lambda}$=1.64$\mathrm{\mu}$m/Br$\mathrm{\gamma}$ vs H$_2$(1-0)S(1)/Br$\mathrm{\gamma}$ and [Fe II]
$\mathrm{\lambda}$=1.25$\mathrm{\mu}$m/Pa$\mathrm{\beta}$ vs [P II]$\mathrm{\lambda}$=1.18$\mathrm{\mu}$m/Pa$\mathrm{\beta}$
diagrams, afforded by the extended IR wavelength coverage offered by the X-shooter data. Unfortunately, the MaNGA wavelength
coverage stops at 
\textasciitilde10000$\mathrm{\AA}$, hence preventing their use for disentangling star formation in these galaxies.

However, we have investigated whether such shocks can potentially mimick
star formation-like line ratios observed by us in galactic outflows by
employing
the MAPPINGS III library of fast radiative shock models \citep{Allen2008}. The predictions of
extensive grids of these models, spanning a wide range of velocities and physical
parameters, have been overlayed on the BPT diagrams and on the R$_{23}$ vs O$_{32}$ diagram, together with the location of the line ratios
observed in the Manga outflows. These various plots are shown in the Appendix. Not unexpectedly,
shock models can reproduce well several line ratios in the LIER region, as well as in the AGN-Seyfert regions. However, although
some shock models do overlap with the loci occupied by star forming regions, the vast majority of the line ratios observed in the star forming outflows
identified by us (blue symbols) are inconsistent with the shock models, further reinforcing the claim that we are indeed observing star formation
inside these outflows.

The additional issue with shocks is that they are much less effective in producing ionizing photons than young, hot stars, hence much less likely
to account for the large nebular luminosities observed in the outflow. Assuming a typical radius of the putative shocked region of $\sim$3~kpc (as
inferred from our maps), by using the equations given in \cite{Allen2008}, focusing on slow shocks
($\rm v_s\sim 100 ~km/s$), which are the ones that may mimic star-forming like line ratios, and assuming the extreme (unrealistic) case of
shocked gas covering the entire 4$\pi$ solid angle with covering factor of unity, we infer that the expected ionizing photon luminosity should
be about $\rm Q_{ion} \approx 10^{52}~s^{-1}$. Assuming that these photons are
entirely absorbed by the gas (by ionizing it), this would result
into a H$\alpha$ luminosity of the outflowing gas of $\sim 10^{40}~erg~s^{-1}$. Much more realistically, the putative shocked gas is characterized by
a much smaller solid angle (outflows, when observed at high angular resolution have typically solid angles less than $\pi$, e.g. Venturi et al. 2018),
with small
covering factor (0.02--0.2, e.g. Baskin \& Laor 2005) and only a fraction of the ionizing photons is likely absorbed by the outflowing clouds,
hence resulting in to an expected H$\alpha$ luminosity one or two orders of
magnitude lower, i.e. $\sim 10^{38}-10^{39}~erg~s^{-1}$. Except for one case,
all outflows that we have classified as star forming have a H$\alpha$ luminosity in the range $10^{40}-10^{41}~erg~s^{-1}$, hence very unlikely
to be produced by shocks.

The combination of these energetics arguments and the location of shock models on the diagnostic diagrams discussed above, provide
strong evidence against any significant role of shocks in the ionization of the outflows that we have identified as hosting star formation.

\begin{figure}
\centerline{\includegraphics[width=9truecm,angle=0]{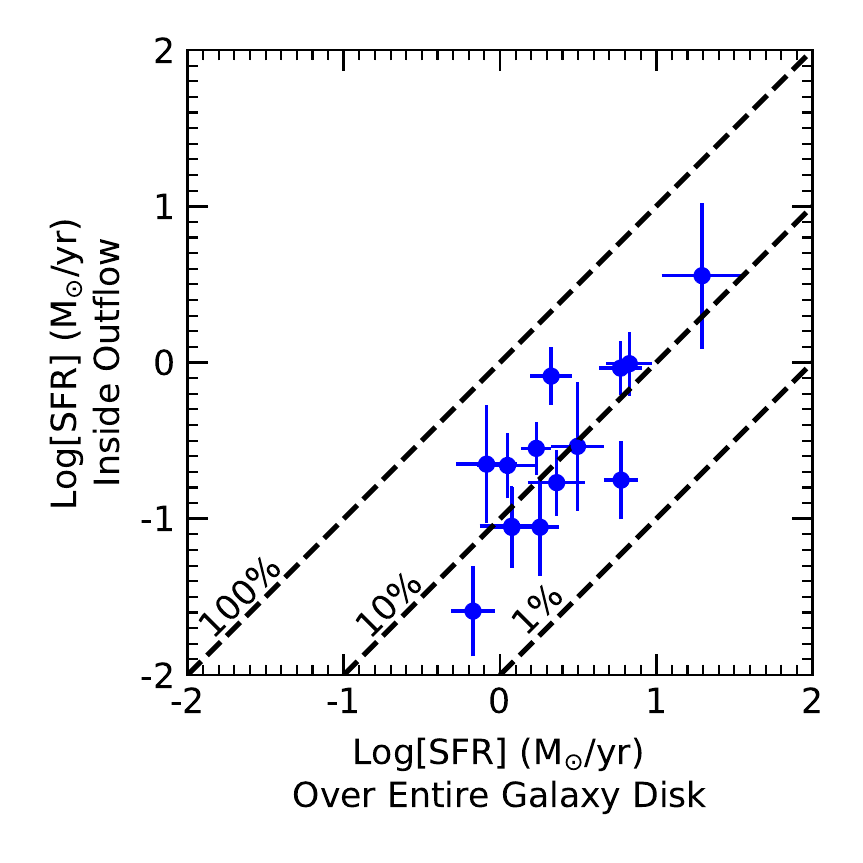}}
\caption{Star formation rate inside the outflow as a function of total star formation rate in the disc of the host galaxy.
}
\label{fig:sfr_relat_total}
\end{figure}

\begin{figure}
\centerline{\includegraphics[width=9truecm,angle=0]{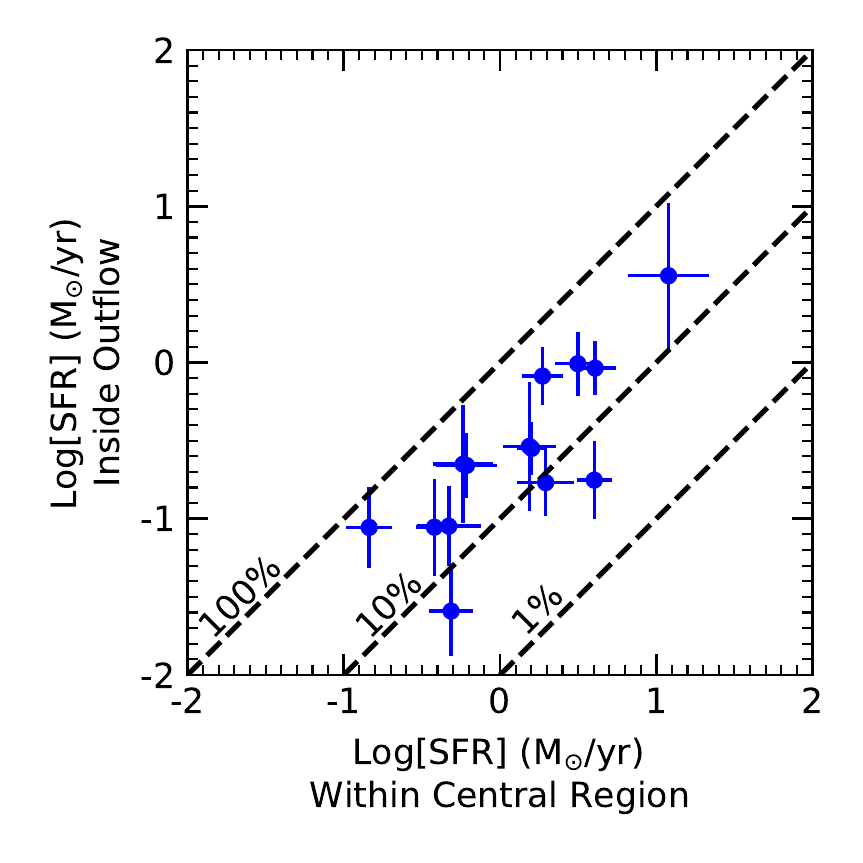}}
\caption{Star formation rate inside the outflow as a function of the star formation rate in the same central projected
area in which the outflow is detected.
}
\label{fig:sfr_relat_central}
\end{figure}

\section{Star formation rate inside outflows} 
\label{sec:sfr_in_outflow}

\subsection{Contribution to the total star formation}
\label{sec:sfr_relat}

For those outflows showing evidence for star formation, we can estimate the star formation
rate by using the broad component of H$\alpha$ as a tracer of star formation \citep{Kennicutt2012}. The
star formation rates inferred in these local outflows are of the order of 0.1--1 $M_{\odot}~yr^{-1}$ (Table \ref{tab:sfr_outf}).
In Figure \ref{fig:sfr_relat_total} we compare the star formation rate inside the outflow with the total star formation
rate of the galaxy, as inferred by the total H$\alpha$ emission (both narrow and broad), integrated across the entire galaxy (excluding
AGN and LIER-like regions).
Star formation in the outflow contributes ``only'' between 5\% and 30\% to the total star formation
rate of these local galaxies (with one extreme case dominating the total SFR).

However, in the spaxels where multiple components are detected (typically within the central few kpc), the star formation inside the
outflow can dominate the local star formation formation rate. This is illustrated in Figure \ref{fig:sfr_relat_central}, where the star
formation rate inside the outflow is compared with the star formation rate in the region in which the outflow is detected (i.e. where a broad
component is detected, which is generally in the central region).

These star formation rates are certainly modest and may be regarded as currently not very relevant for galaxy evolution.
However, one should also take into account that these are relatively mild outflows, while star formation may be much
more prominent inside the massive outflows driven by massive star forming galaxies at high redshifts or powerful distant quasars.
To investigate this possibility in the following we explore the scaling relation between star formation inside the
outflow and mass outflow rate.

\subsection{Ionized outflow rate}
\label{sec:outfl_rate}

Unfortunately we only have information on the ionized component of the outflowing gas, therefore we can only obtain
information on the ionized gas outflow rate. The ionized phase generally accounts for a small fraction of the total gas
content in outflows \citep{Carniani2015a,Fiore2017,Fluetsch2018}, however it can be used as a proxy of the global outflow rate.

We have adopted a method similar to the one described by \cite{Carniani2015a} and \cite{Cicone2015}.
More specifically, for each point where we detect an outflow component we estimate the contribution
to the outflow rate as the mass of ionized gas at that point divided by the dynamical time from the centre:

$$\rm \dot{M}_{outf-ion}=\frac{{M}_{outf-ion}}{\tau _{dyn}}=\frac{{M}_{outf-ion}~v_{outfl.}}{R_{outfl}}$$

where $\rm v_{outfl}$ is the outflow velocity
at that location, which we define as $\rm |v|+FWHM/2$ of the broad component
($\rm |v|$ being the absolute velocity of the centre of the broad component
tracing
the outflow, relative to the galaxy rest frame), $\rm R_{outfl.}$
is the distance of the specific location to the galaxy centre (within the central PSF region this is taken as the size of the PSF,
which actually gives a lower limit to the contribution to the outflow rate in this region) and $\rm M_{outfl-ion}$,
is the mass of ionized gas at that location (spaxel). Projection effects, resulting from the direction of the outflowing gas relative to the
line of sight, affect both the observed velocity and the observed radius. However, it can be easily shown \citep{Cicone2015}
that the correction factor, averaged over the entire 4$\pi$ solid angle, is one; therefore, statistically,
projection effects do not plague results, although they certainly introduce scatter. Assuming $\rm T_e \sim 10^4~K$,
the mass of ionized gas in the outflow at that location is given by the luminosity of the [OIII]5007
emission at the same location through the following relation:

$$\rm M_{out-ion,[OIII]} = 2~10^8 M_{\odot}~10^{-[O/H]} \left( \frac{L_{[OIII]}}{10^{44}~erg~s^{-1}}\right)
\left( \frac{n_e}{200~cm^{-3}}\right) ^{-1}$$

where [O/H] is the oxygen abundance relative to solar and $\rm n_e$ is the electron density in the outflow, which we have
assumed to be $\rm 200~cm^{-3}$, which is intermediate between the results
obtained by \cite{Perna2017b} (for AGN-driven outflows)
and the result obtained by \cite{Genzel2014} (for outflow in massive star forming galaxies).
In a future work (Gallagher et al., in prep.) we will refine the outflow rates by taking the gas density inferred directly from the sulphur doublet at each
location of each outflow.

Alternatively, the mass of the ionized outflow can be inferred from the broad component of H$\alpha$ from the following equation

$$\rm M_{out-ion,H\alpha} = 1.5~10^9 M_{\odot} \left( \frac{L_{H\alpha}}{10^{44}~erg~s^{-1}}\right)
\left( \frac{n_e}{200~cm^{-3}}\right) ^{-1}$$

As already pointed out by \cite{Carniani2015a}, if assuming solar metallicity, the outflow mass inferred from [OIII]
broad component gives an outflow rate which is about a factor of $\sim 3-5$ lower than inferred from H$\alpha$ or H$\beta$.
This is either because the gas metallicity may be lower than solar or, more likely, because [OIII] does not account
for the contribution from lower ionization stages of oxygen. 

In either cases, the total outflow rate was then obtained by summing the individual contributions to the outflow rate of all individual spaxels.

\begin{figure}
\centerline{\includegraphics[width=9truecm,angle=0]{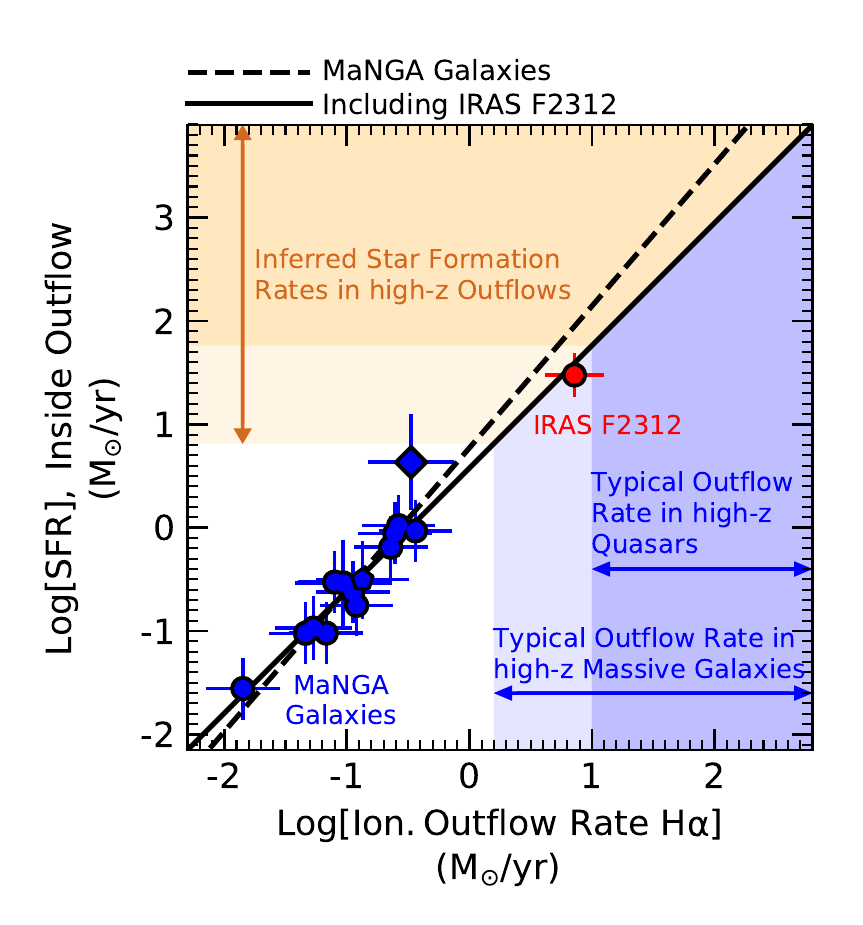}}
\caption{
Star formation rate inside the outflow as a function of ionized gas mass outflow rate inferred from the H$\alpha$.
Blue symbols are star forming outflows in local MaNGA galaxies studied in this paper. The red symbol indicates the
star forming outflow previously detected in IRAS F23128-5919 \citep{Maiolino2017}.
The dashed line shows the fit to the MaNGA outflows presented in this paper (blue points). The solid line is the linear
fit including IRAS F23128-5919 (Eq. \ref{eq:sfr_vs_outf_ha}).
The blue shaded regions indicate the ionized outflow rate inferred for high-z massive galaxies and quasars.
The orange shaded regions indicates
the star formation rate inside these distant galactic outflows if they follow the same correlation observed locally.
}
\label{fig:sfr_vs_outf_ha}
\end{figure}

\begin{figure}
\centerline{\includegraphics[width=9truecm,angle=0]{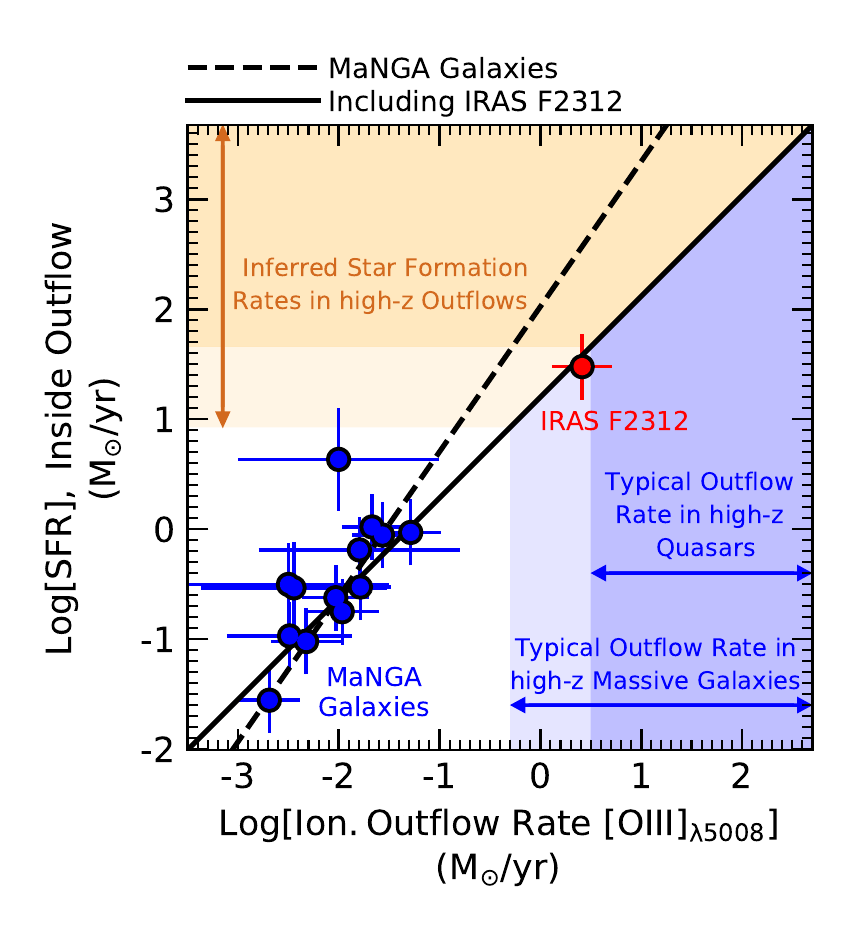}}
\caption{
As Figure \ref{fig:sfr_vs_outf_ha}, but where the ionized outflow rate is now inferred from the broad component of the [OIII] line.
}
\label{fig:sfr_vs_outf_oiii}
\end{figure}

\subsection{Scaling relations and implications for high redshift galaxies}
\label{sec:scaling}

We have explored the connection between outflow rate and star formation in the outflow.

We have initially focused the outflow rate inferred from the H$\alpha$ broad component, because less subject to ionization and metallicity
corrections.
For those galaxies whose outflow is dominated by star formation in terms of BPT diagnostics, Figure \ref{fig:sfr_vs_outf_ha} shows
the star formation rate inside the outflow as a function of the ionized outflow rate estimated from the broad component of H$\alpha$.
Blue symbols are star forming outflows discovered in this paper, while the red symbol is the star formation previously discovered inside the outflow
of the galaxy IRAS F23128-5919 \citep{Maiolino2017}. The data reveals a clear correlation between the two quantities. To perform
linear fits to the data, accounting for the errors in both the outflow rate and star formation rate, we used the the
\texttt{scipy} Orthogonal Distance Regression (ODR) package. The errors in these quantities were calculated through rigorous error 
propagation, though we elected to maintain a minimum error or 0.3 dex to account for previous considerations, such as assuming density
values of $\rm 200~cm^{-3}$ and LOSVD errors being the formal errors on the fit returned from \texttt{mpfit}. We began by performing a fit
to the MaNGA outflows presented in this paper alone. This fit is shown by the dashed line, and extrapolation of this fit is nicely consistent
with the star formation observed in the outflow of IRAS F23128-5919. This fit takes the form:

\begin{equation}
\rm \log{\left( SFR_{outfl}\right)} = 1.37(\pm 0.14) ~\log{\left( \dot{M}_{outfl-ion,H\alpha}\right)}+0.77(\pm 0.14)
\label{eq:sfr_vs_outf_ha_MaNGA}
\end{equation}

Including the latter galaxy in a secondary fit results in the fit shown by the solid 
line, which is only slightly different from the fit to the MaNGA galaxies alone, and takes the form:

\begin{equation}
\rm \log{\left( SFR_{outfl}\right)} = 1.19(\pm 0.07) ~\log{\left( \dot{M}_{outfl-ion,H\alpha}\right)}+0.58(\pm 0.07)
\label{eq:sfr_vs_outf_ha}
\end{equation}

We recall again that generally the ionized outflow rate is a small fraction of the total outflow rate (especially relative to the molecular
phase). The blue shaded regions indicate the ionized outflow rate inferred for high-z massive galaxies and quasars,
as inferred from H$\alpha$ or H$\beta$ data \citep{Carniani2015a,Genzel2014,Fiore2017}. The orange shaded regions indicates
the star formation rate inside these distant galactic outflows if they follow the same correlation observed locally, implying
star formation rates in outflows ranging from $\rm \sim 10~M_{\odot}~yr^{-1}$ up to $\rm \sim 1000~M_{\odot}~yr^{-1}$, which would
certainly contribute significantly to the galaxy formation, and to their spheroidal component in particular.

Such high star formation rates in distant galactic outflows may have been missed by past observing campaigns and in section
\ref{sec:agn_dominance} we will discuss the possible reasons for this.

One potential problem of the diagram in Figure \ref{fig:sfr_vs_outf_ha} is that the broad H$\alpha$ flux is a quantity used to measure
both the ionized outflow rate (together with the additional information on the outflow velocity and distribution) and to estimate the
SFR in the outflow, which may induce a spurious correlation. Thefore, in order to test whether the correlation persist even by
adopting different tracers that use independent quantities, we have considered the outflow rate estimated through the [OIII] line.
As discussed above, [OIII] is more unreliable as an outflow tracer (because of its dependence on the metallicity and on the
ionization degree) and tends to underestimate the outflow rate, however it has the advantage that it is completely independent of the H$\alpha$ flux
used to estimate the star formation rate.
Figure \ref{fig:sfr_vs_outf_oiii} shows the star formation in the outflow as a function of the ionized outflow rate as inferred from the
broad component of the [OIII] line. Again, two fits were performed: firstly to the MaNGA outflows alone, and secondly to the MaNGA outflows including IRAS F23128-5919 in the fit. The two quantities are still clearly correlated, although the scatter is slightly larger than in
Figure \ref{fig:sfr_vs_outf_ha}. In this case the fit to the MaNGA outflows alone (dashed line) is less consistent with the outflow in IRAS 
F23128-5919, and takes the form:

\begin{equation}
\rm \log{\left( SFR_{outfl}\right)} =  1.32(\pm 0.18)~\log{\left( \dot{M}_{outfl-ion,[OIII]}\right)}+2.02(\pm 0.37)
\label{eq:sfr_vs_outf_oiii_MaNGA}
\end{equation}

However, the larger scatter in the data makes the extrapolation of the MaNGA data less reliable. The solid line shows the linear fit 
including both the MaNGA outflows and the outflow in IRAS F23128-5919, which is given by the following equation

\begin{equation}
\rm \log{\left( SFR_{outfl}\right)} =  0.92(\pm 0.06)~\log{\left( \dot{M}_{outfl-ion,[OIII]}\right)}+1.2(\pm 0.1)
\label{eq:sfr_vs_outf_oiii}
\end{equation}

The slope, close to unity, is very similar to the slope obtained in Eq. \ref{eq:sfr_vs_outf_ha} when using H$\alpha$ as a tracer of the
star formation rate. The intercept is obviously different, as a consequence of [OIII] underestimating the ionized outflow rate.
The blue and orange shaded regions in Figure \ref{fig:sfr_vs_outf_oiii} have the same meaning as in Figure \ref{fig:sfr_vs_outf_ha}
(in this case the range of outflow rates, based on [OIII], in high-z galaxies is more incomplete and we have simply assumed the
ranges inferred from H$\alpha$ an H$\beta$ and shifted by a factor of 0.5~dex, as discussed in Sec. \ref{sec:outfl_rate}).
As in the previous case, the extrapolation to the outflow rates of high redshift galaxies implies high star formation rates
in the outflows of distant galaxies.

We shall recall that stars obviously form out of the molecular phase of galactic outflows, therefore it would certainly be more
appropriate to exploit scaling relations involving the molecular outflow rate. Unfortunately, information on the molecular outflow rate
is not yet available for these star forming outflows (although ALMA observations are ongoing for some of these systems). However, it should 
be noted that \cite{Fluetsch2018} have found that the molecular-to-ionized outflow rate increases with AGN luminosity \citep[an opposite trend
with respect to what found by ][ who however use disjoint samples to probe the ionized and molecular phases, hence potentially subject to differential
selection effects]{Fiore2017}; therefore, in powerful outflows driven by luminous AGNs the fraction of stars forming in outflows may potentially
be even larger than that inferred from the scaling relations with the ionized component of outflows given by Equations \ref{eq:sfr_vs_outf_ha} and \ref{eq:sfr_vs_outf_oiii}.

\section{Distribution of galaxies with outflows on the SFR-M$_*$ diagram}
\label{sec:MS}

\begin{figure}
\centerline{\includegraphics[width=9truecm,angle=0]{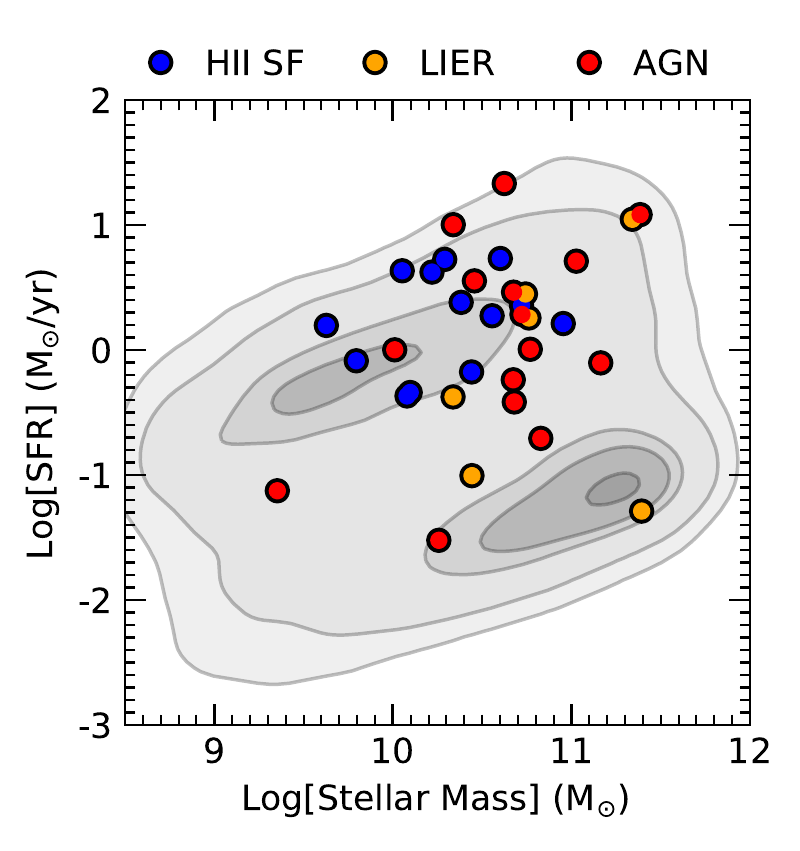}}
\caption{
Star formation rate versus stellar mass diagram. The background contour plot shows the distribution of all 2,800 galaxies in the MaNGA sample.
The symbols show the distribution of galaxies with outflows presented in this study, i.e. those for which the outflows can be analysed in the BPT diagram.
The BPT classification of the outflow is the same as in the [SII]-BPT diagram of Fig.\ref{fig:bpt_total}.
Most outflows, especially those hosting star formation inside the outflow, are located around the massive end of the Main Sequence, or slightly above it.
}
\label{fig:MS}
\end{figure}

The distribution of galaxies on the SFR-M$_{\rm star}$ diagram is often used to identify their properties. Most star forming galaxies are
distributed along the so-called ``Main Sequence'',
a tight relation in which the SFR is nearly proportional to the stellar mass \citep{Renzini2015}. This
is thought to represent the sequence along which galaxies form stars through secular, smooth evolution. Passive galaxies have SFRs well
below the ``Main Sequence'', while galaxies distributed between the Main Sequence and the Passive population (in the region that is often
referred to as ``green valley'') are believed to be in the process of being quenched or being ``rejuvenated''.
The distribution of the 2,800
MaNGA galaxies (our parent sample) on the SFR-M$_{star}$ diagram is shown with grey contours in Figure \ref{fig:MS}
(SFR and M$_{star}$ are obtained from the DR7
catalogue\footnote{https://wwwmpa.mpa-garching.mpg.de/SDSS/DR7/}).
The Main Sequence in this sample
is less populated than in other samples (such as, for instance, SDSS-DR8) primarily as a consequence of the MaNGA selection criterion of
having a flat distribution in stellar mass. The MaNGA galaxies with outflows investigated in this paper, i.e. those for which the
outflow can be studied in terms of BPT diagram, are shown with symbols, which are color-coded according to the [SII]-BPT classification
of the gas in the outflow as in Figure \ref{fig:bpt_total}. Most of the galaxies with outflow are distributed around the massive end of the Main Sequence.
Outflows hosting star formation (blue symbols) are along the Main Sequence or slightly above it. There are some outflows among passive
and green-valley galaxies, but only with AGN and LIER-like excitation. The lack of outflows in low mass galaxies may primarily result
from selection effects, as these tend to be fainter and are likely characterized by milder outflows (either SF-driven or AGN-driven),
below our detection threshold. A more detailed analysis of the statistical properties of outflows will be discussed in another paper.
However, the finding that outflows hosting star formation are located primarily on, or slightly above, the massive end of the main
sequence supports the scenario in which this star formation mode may contribute to the population of stars in the bulge; although the
overall contribution to the bulge population at this late epoch is minor, the contribution of this mechanism to the evolution of the bulge may have been
significant at high-z, in the early phase of galaxy formation. This result is in line with the recent finding that galaxies
above the Main Sequence have bluer bulges, which may still be in the process of forming a fraction of their stars \citep{Morselli2017}.

We finally mention that, both morphologically and kinematically, nearly all galaxies showing star formation inside outflows
appear to be isolated systems, typically regular galaxy disc. Only two of these galaxies are interacting systems, which are marked with an asterix
in Table \ref{tab:z_mass}.

\section{Why has star formation in outflows been elusive?}
\label{sec:agn_dominance}

If star formation in outflows is as common as inferred in Sec. \ref{sec:sf_in_outflow} and can reach star formation rates as high
as inferred in Sec.\ref{sec:scaling} in high-z galaxies, why has this phenomenon remained elusive so far?

One of the primary reasons is that different excitation mechanisms do not have the same weight in affecting the location of galaxies or
galactic regions on the BPT diagrams. In particular,
in this section we show that even if prominent star formation is taking place inside the outflow,
the presence of an even weak AGN photoionizing the gas within the outflow generally dominates these diagnostic diagrams,
even in the scenarios with the most conservative assumptions, implying that the fraction of star forming outflows inferred in
section \ref{sec:sf_in_outflow} is actually a lower limit.

To show this, we make two simplifying assumptions: firstly, we assume that the BH accretion and star formation in the spheroidal component of
the galaxy evolve along the local $\rm M_{BH}-M_{spheroid}$ relation; secondly, we take the most extreme (and most conservative)
case possible, wherein we assume that all stars in the spheroid form through star formation within the outflow.

This implies that

\begin{equation}
\rm \dot{M}_{BH} \approx 5~10^{-3}~SFR_{spheroid}
\label{eq:BHacc_sfr}
\end{equation}

\citep{Kormendy2013a}.
Assuming a radiative efficiency of 0.1 for black hole accretion,
the relationship between H$\alpha$ emission (in the NLR) and X-ray luminosity for type 2 AGNs given in \cite{Ueda2015},
and the X-ray to bolometric AGN luminosity correction given in \cite{Marconi2004}
(here we take the bolometric correction appropriate for the typical AGN luminosity considered by \citet{Ueda2015}
implying $\rm L_{bol}/L_{(2-10 keV)} = 20$), we get

\begin{equation}
\rm L_{H\alpha}(AGN)[erg/s] = 2~10^{42}~\dot{M}_{BH} [M_{\odot}/yr]
\label{eq:Ha_BHacc}
\end{equation}

By also using the relation between L(H$\alpha$) and SFR given in \cite{Kennicutt2012}, in combination with Eqs.\ref{eq:BHacc_sfr}--\ref{eq:Ha_BHacc}, we obtain

\begin{equation}
\rm L_{H\alpha}(AGN) \approx 5~ L_{H\alpha}(SFR)
\label{eq:Ha_AGN_SFR}
\end{equation}

and, of course, the same relation applies to H$\beta$. Moreover, for most local star forming galaxies typically
$\rm L_{[OIII]}(SFR)\le L_{H\beta}(SFR)$, whilst for AGNs $\rm L_{[OIII]}(AGN)\approx 10~L_{H\beta}(AGN)$;
as a consequence, together with Eq.\ref{eq:Ha_AGN_SFR}, this implies that the BPT diagnostics will be totally skewed towards the AGN region on their Y-axis.
Similarly, given that [NII], [SII] and [OI] are all more luminous in AGNs than in SF galaxies, together with Eq.\ref{eq:Ha_AGN_SFR}, this implies that the BPT diagnostics
will be all skewed towards the AGN region also on the X-axis. Therefore, even in the most conservative scenario assumed above, in which
the formation of the spheroidal component of the galaxy results entirely from star formation in the outflow,
implying prominent star formation in the outflow, such star formation may go entirely undetected in the BPT diagrams
as the excitation diagnostics are totally dominated by AGN excitation.

This issue has been clearly directly demonstrated in detailed spatially resolved cases in which star forming clumps have remained elusive because
diagnostics have been dominated by the photoionization of an even weak AGN \citep{Santoro2016}.

One consideration that complicates things further is the presence of shocks.
In these cases, the detection of SF is even more difficult as they are expected to move the diagnostics towards the LIER regions on the diagrams.

Summarizing, the lack of star formation-like BPT diagnostics in outflows does not necessarily exclude the presence
of star formation inside the outflow, as such outflows may still host prominent star formation but may struggle
to be detected as other sources of excitation may be dominating the diagnostics.
Star formation inside outflows becomes easier to detect when the AGN has faded away, and when shocks are weak.

The implications in the context of this paper is that the fraction of outflows with star formation may be higher than what
estimated in Sect. \ref{sec:sf_in_outflow}. 

In integrated, spatially unresolved (or poorly resolved) spectra the light-weighting further enhance the bright, central AGN-dominated
emission (especially in the case of powerful quasars)
relative to weaker star forming regions. This issue is obviously more
problematic at high redshift due to the lower spatial resolution.
However, dispite these difficulties, current data already show some indications of star formation in high redshift outflows.
\cite{Genzel2014} investigated the outflows of five high-z galaxies and two of them have diagnostic
ratios consistent with star formation inside the outflow.

The finding that at high redshift star forming regions tend to shift towards the AGN locus on the BPT diagrams \citep[e.g.][]{Strom2017a}
is a further complication which may have led the classification of some
outflows as AGN-dominated, while their BPT-ratios may actually be
associated with star formation inside the outflow.

Another problem is that often only the [NII]-BPT diagram is used to classify outflows. As discussed in \cite{Strom2017b} and
\cite{Masters2016} the [NII]-BPT is much more subject to changes in the ionization parameters and nitrogen abundance
relative to the other two BPT diagrams. In Sect. \ref{sec:sf_in_outflow} we have seen that star forming outflows are clear identified
in the [SII]- and [OII]-BPT diagrams, while in the [NII]-BPT diagram they are located primarily in the ``composite'' region.
Therefore, studies based only on the [NII]-BPT diagnostics
may have mis-identified the excitation of galactic outflows. For instance, \cite{Leung2017} classify the outflows of 13 AGNs at z$\sim$2
only based on the [NII]-BPT diagram; about half of these outflows are located in the ``composite'' region of the [NII]-BPT diagram,
hence these outflows could be hosting star formation, similarly to the outflows investigated in this paper.

Another problem is that in many spectroscopic surveys outflows are searched through blue wings of the [OIII]5007 line. Because of the
dominance of this line in AGN-driven outflows, as discussed above, many of these surveys have certainly biased their search toward
outflows whose diagnostics are AGN-dominated.

Finally, with the growing interest in studying local galaxy outflows in much greater details thanks to improved, high sensitivity and wider
field of view integral field spectrometers (e.g MUSE at the VLT), alongside evidence for star formation being found in additional individual
outflows studied in detail \citep[Venturi et al., in prep.; Fluetsch et al., in prep.,][]{Maiolino2017}, we expect to see many highly detailed
and successful studies of star formation within outflows within the years to come.


\section{Implications for galaxy evolution}
\label{sec:implications}

Stars formed in galactic outflows have completely different dynamical properties relative to the stars forming in galactic discs.
Indeed, as soon as stars form in the outflow they only respond to gravity, hence moving ballistically. If the velocity at the
time of formation inside the outflow exceeds the escape velocity then the newly formed stars will leave the galaxy and
disperse in the intergalactic medium (IGM). If the velocity is lower than the escape velocity then the newly formed stars will
be bound; they will be rapidly decelerated and will start oscillating around the galactic center.
The latter is also one of the reasons why it is difficult to identify the young stars formed in the outflows, since
as soon as they form they decouple from the outflow and their velocity is quickly reduced \citep[within a few Myr,][]{Maiolino2017}
becoming kinematically difficult to distinguish from the stars in the disc.

In the case of the stars formed in the outflows that are gravitationally bound, their resulting orbits
may predominantly be radial. However, as shown in \cite{Maiolino2017} for the specific case of IRAS F23128-5919, the gravitational
interaction with the galactic disc, or any other non-spherically symmetric potential (stellar bar, flyby galaxies) will introduce
a tangential component to the stellar velocities, hence randomizing their orbits. Moreover if the expelled gas was initially
in rotation around the galactic centre, then the resulting stars formed in the outflow star can also preserve some minor
rotation component.

What is the primary fate of stars formed in the outflow has still to be investigated properly in a statistical way. Models
expect both scenarios, i.e. escaping and gravitationally bound,
depending on the nature and properties of the outflow \citep{Zubovas2013a}. In the case of the
star forming outflow in IRAS F23128-5919, which has been studied in great detail, \cite{Maiolino2017} has shown that most of
the stars born in the outflow are gravitationally bound. A detailed study of both the MaNGA galaxies shown here (as well as those for
which a BPT analysis has not been possible) is postponed to a later paper, however a rough analysis of the kinematics and of the galaxy dynamics suggest a broad
range of possibilities, from stars on close, gravitationally bound orbits to stars escaping the galaxy potential.

In either case, the fact that star formation in outflows has been found by us to be so common implies that this newly discovered phenonenon
may have major implications for galaxy evolution and even result into a paradigm change in some specific areas.
In the following we discuss some of the potential implications of star formation in outflows for galaxy evolution
and to explain some of the galactic properties.

\subsection{Formation of the spheroidal component of galaxies}

If stars formed in outflows are gravitationally bound \citep[this seems to be the case in most cases, as the broad component velocity dispersion
is generally comparable or slightly lower than the escape velocity, as expected by some models, ][]{Zubovas2013}, then they can contribute significantly to
the formation and evolution of the spheroidal component of galaxies, i.e. bulge, halo, and even contribute
to the formation of elliptical galaxies.
Indeed, \cite{Silk2013} have proposed that such positive feedback mechanism
may explain some of the extreme star forming galaxies at high redshift, which are thought to be projenitors
of local ellipticals. \cite{Ishibashi2013} have even suggested that star formation in AGN-driven outflows can contribute
to the observed size evolution of ellitpical galaxies across the cosmic epochs. \cite{Wang2018} have also suggested that
the double-shells often observed in elliptical galaxies, typically ascribed to merging events, can actually be the
result of star formation in outflows.

The integrated spectra of local elliptical galaxies does not seem to show evidence a large fraction of stars on radial
orbits \citep{Cappellari2007}. However, the study of \cite{Cappellari2007} of 24 local elliptical galaxies results into a radial anisotropy
parameter $\beta _r >0$ (i.e. preference for radial orbits) in $\sim 2/3$ of their sample. The kinematics of individual stars
would be needed to properly assess the fraction of the radial orbits.

For galactic bulges and galactic halos, for which star formation within outflows is likely more important,
a similar analysis on the distribution of radial and tangential orbits has not
yet been performed to our knowledge. It is however interesting to note that recent studies of the stellar velocity
distribution in the 5~kpc around the Sun, based on Gaia data, have revealed a significant population of stars
with highly radial galactocentric orbits \citep{Myeong2018,Belokurov2018}. The second Gaia data release is expected to
further expand the characterization and detection of stars on radial orbits.

Within this context star formation in outflows has also been invoked to explain the population of
Hypervelocity stars in the Galactic halo \citep{Wang2018,Silk2012,Zubovas2013a}.

\subsection{BH--galaxy correlations}

Clearly stars formed in AGN-driven outflows can potentially directly produce the
$\rm M_{BH}-M_{spheroid}$ relation.
In particular, given the correlation between outflow rate and AGN luminosity (i.e. BH accretion rate), obtained
by various observational works \citep{Fluetsch2018,Fiore2017,Cicone2014}, and assuming the correlation between outflow
rate and star formation in the outflow found in Sect. \ref{sec:scaling} (Figs. \ref{fig:sfr_vs_outf_ha}-\ref{fig:sfr_vs_outf_oiii}) holds true for 
all AGN driven outflows,
it is clear that the combination of the two can result into a correlation between black hole mass
and stars formed in the outflows.

Quantifying the expected relation is yet not simple, as different authors have
found different scaling relations between ionized outflows and AGN luminosity \citep{Fiore2017,Fluetsch2018}.
However, if we consider the compliation by \cite{Fiore2017} (where the correlation between ionized outflow rate and AGN luminosity
is slightly superlinear), and assume an AGN radiative efficiency of 0.1, then the relation between
AGN luminosity and ionized outflow rate translates into

\begin{equation}
\rm \log{\left(\dot{M}_{outf-ion}\right)} = 0.85+1.29 \log{\left(\dot{M}_{BH}\right)}
\label{eq:outf_bhacc}
\end{equation}

where both the ionized outflow rate and black hole accretion rate are in units of $\rm M_{\odot}~yr^{-1}$, and where we have corrected the outflow rates estimated
by \cite{Fiore2017} (specifically by a factor of three) to account for the different assumptions to
derive the outflow rate in our work.
Combining Eq.\ref{eq:outf_bhacc} with the relation between ionized outflow rate and star formation in the outflow,
i.e. Eq.\ref{eq:sfr_vs_outf_ha} gives

\begin{equation}
\rm  SFR_{outf-ion} = 15 \left(\dot{M}_{BH}\right)^{1.56}
\label{eq:sfrout_bhacc}
\end{equation}

were $\rm  SFR_{outf-ion}$ is the star formation rate inside the outflow, again in units of $\rm M_{\odot}~yr^{-1}$.
The relation is superlinear. At low/intermediate accretion rates ($\rm
\dot{M}_{BH} < 5 M_{\odot}~yr^{-1}$)
the relation gives a
$\rm SFR/\dot{M}_{BH}$ ratio lower than the observed local
bulge-to-black hole mass ratio, i.e.
$\rm M_{bulge}/M_{BH}\sim 200$ \citep{Kormendy2013a}, while at high accretion
rates ($\rm \dot{M}_{BH} > 5 M_{\odot}~yr^{-1}$) the implied $\rm SFR/\dot{M}_{BH}$
ratio is higher than the local bulge-to-black hole mass ratio. This behaviour
in two different regimes
can potentially explain the very low $\rm M_{star}/M_{BH}\sim 30$ observed at high
redshift \citep[e.g.][]{Wang2013,Decarli2018}, which can be reached
through the first (low accretion rate)
regime, in which the black hole forms faster than the stellar population
(relative to the local relation), while later (in the high accretion rate regime)
the implied higher star formation in the outflow can enable the galaxy to build up
quickly enough stellar mass to reach the local relation.

Of course, due to the relations being non linear and the correlations not yet well established, more work is needed to properly
assess exactly whether the BH-galaxy correlations and their evolution can be fully explained through star formation in outflows or not.

It is finally interesting to note that \cite{Fiore2017} and \cite{Bischetti2018} have identified a very steep
relationship between outflow velocity and AGN luminosity. More specifically,
the outflow velocity (combining molecular and ionized outflows) scales with AGN luminosity, i.e. black hole accretion
rate, as 
\begin{equation}
\rm L_{AGN}\propto v_{outfl}^{4.6}
\end{equation}

The slope of this correlation is comparable to the slope of the correlation between black hole
mass and stellar velocity dispersion:
\begin{equation}
\rm {M}_{BH}\propto  \sigma _v ^{4.5-5}
\end{equation}

\citep[e.g.][]{Shankar2016}. Therefore, if during the active outflow phases black holes are accreting close to the
Eddington limit (hence $\rm L_{AGN}\propto M_{BH}$), as expected by many outflow models \citep[][]{King2015},
and stars form inside such AGN-driven outflows, then
the stellar velocity dispersion of the resulting spheroidal stellar component
would have a velocity dispersion consistent with that of the $\rm M_{BH}-\sigma$ relation.

\subsection{Extragalactic supernovae, CGM/IGM in-situ enrichment and halo heating}

The population of stars formed in galactic outflows can result in supernovae exploding on large orbits, outside the galaxy.
As discussed in the previous sections most of the stars formed in the outflow are rapidly decelerated (wihtin a few Myr) and
will quickly fall back onto the galaxy, hence a significant fraction of the associated supernovae explode close to the galaxy.
However, a fraction of stars in the outflow may have larger orbits or even escape the galaxy, as expected in models \citep{Zubovas2013a}
and as observed \citep{Maiolino2017}. The resulting supernovae should appear as supernovae exploding outside galaxies.

Search for supernovae
tend to be biased toward supernovae exploding inside galaxies, as monitoring and followup is typically focussed on galaxies.
However, hostless or intergalactic SNe have been observed in clusters \citep{Gal-Yam2003,Graham2015,Gupta2016}.

It should be reminded that locally star formation in outflows is a relatively mild phenomenon, hence supernovae outside
galaxies are not expected to be very common. However, at high-z, where this phenomenon is expected to be much more prominent,
supernovae exploding in the Circum-Galactic Medium (CGM) or even in the Inter-Galactic Medium (IGM) should have been much more common.
Such extragalactic SNe would have directly enriched, {\it in-situ}, the CGM and IGM, relaxing the need of metals to be expelled
from galaxies through winds.

An additional implication is that SNe in the halo would be expanding in a very low density environment, hence releasing most of the
energy to the halo \citep[in contrast to galactic SNe, which explode in dense environments hence losing most of their energy through
radiative losses, ][]{Walch2015}. As a consequence, SNe outside the galaxy can be very effective in heating the halo, hence supperssing
the accretion of cold gas onto the galaxy, resulting in a delayed (``preventive'') feedback, which may result in quenching
star formation as a consequence of starvation. Therefore, while galactic outflows have a positive initial effect, resulting in new
stars born in the outflow, eventually they may actually have a net negative feedback effect on the entire galaxy through halo
heating and galaxy starvation.


\subsection{Reionization of the Universe}

As discussed above, as soon as stars form
in the outflow they only respond to gravity, hence they quickly decouple from the outflowing clouds (which are instead
also subject to radiation pressure, ram pressure and other fluidodynamic effects). As a consequence, the massive young, hot
stars formed in the outflow will have very high escape fraction of ionizing photons. Therefore, if this phenomenon is
taking place in the early Universe, star formation in outflows can greatly contribute to the reinoization of the Universe.
Indeed, one of the main problems of current theories is that the ionizing photons escape fraction typically observed in galaxies
is very low (less than 5\%-10\%), making it diffucult for models to account for the reionization of the Universe.
The large escape fraction expected for stars formed in outflows may help to tackle this long standing issue.

\section{Conclusions} 

Several models have proposed that massive galactic outflows may form stars
inside them, with potentially far reaching implications as stars formed inside outflows
would have completely different kinematic properties than those formed in galactic discs.
The detection of large amounts of molecular gas, both dense and clumpy, in galactic outflows
does indeed support the scenario that galactic outflows should form stars. However,
direct evidence of star formation inside galactic outflows was so far found only
in a single galaxy, leaving unclear whether this is a rare phenomenon or
common to galactic outflows.

In order to address this issue, we have analysed integral field spectroscopic data of 2,800
local galaxies in the MaNGA DR2. Through a detailed spectral analysis
we have identified a sample of 37 galaxies that
show clear evidence for outflows and whose excitation mechanism can
be investigated through the detection of multiple nebular lines, which
can be used to trace the excitation of outflowing gas through the BPT diagnostic diagrams.

We have obtained the following results:

\begin{itemize}

\item About 30\% of the outflows have BPT diagnostic diagrams consistent
with being star forming, i.e. the nebular lines in the outflow are primarily
excitated by the UV radiation of young hot stars. Moreover, we find
that half of the outflows show at least some fraction of them ($>$10\%) being excited
by star formation. These results highlight that star formation is common
in a large fraction of galactic outflows. 

\item The analysis of the ionization parameter in the star forming outflows, which is
indistinguishable from (or even higher than) normal star forming regions,
confirms that the outflowing
gas is not photoionized by the UV radiation coming from the underlying
galactic discs, but must be photoionized by {\it in-situ} (i.e. within
the outflow) star formation.

\item In these local galaxies the star formation inside the outflow generally
accounts only for about 5\%--30\% of the total star formation in the galaxy.
However, star formation inside the outflow can contribute significantly, or
even dominate star formation in the central few kpc.

\item We find that, for those galaxies whose outflow is dominated by star
formation, the star formation rate inside the outflows correlates
with the ionized outflow rate. If extrapolated to the outflow rates observed
in distant massive galaxies and quasars, then the implied star formation
rate inside those outflows can be up to several 100~$\rm M_{\odot}~yr^{-1}$,
hence potentially contributing significantly to the evolution of galaxies.
We suggest that evidence for star formation in the outflows of distant
galaxies may already be present.

\item We show that galaxies with star formation inside the outflow
are primarily distributed along the ``Main Sequence'' (or, in some cases, slightly above it) and are typically isolated,
regular disc galaxies.

\item We discuss the possible reasons why star formation may have remained
elusive until recently. We show that even if prominent star formation is
present inside the outflow, the presence of an even faint AGN generally
dominates the diagnostics. This may have precluded the detection of
star formation in many outflows, especially in integrated spectra or
with poor spatial resolution. We also point out that outflows identified
through the [OIII]5007 line may have biased the samples against those
dominated by star formation. Moreover, the use of the [NII]-BPT diagnostic
diagram may also have resulted in improperly
classifying the nature of some outflows.

\item As soon as stars form inside galactic outflows they react only
to gravity, they decouple from the gaseous outflow and move ballistically. Stars formed in galactic outflows
may be either escape the galaxy, or remain gravitationally bound.
As our study has revealed that star formation inside galactic outflows
is a relatively common phenomenon, we have discussed the implications of such
widespread production of stars with kinematic properties completely different
from those formed in galactic discs.
Specifically:

	\begin{itemize}

	\item Star formation inside outflows can potentially contribute
	significantly to the formation of the spheroidal component of galaxies
	(bulge, halo, and possibly also part of ellitpical galaxies).
	Within this context stars formed inside galactic outflows can explain
	the population of stars on radial orbits recently discovered in our
	Galaxy, as well as the population of hyper-velocity stars.
	The same phenomenon can explain she double-shells observed in
	ellitpical galaxies.

	\item This phenomenon can contribute to establish the correlations between
	black holes and their host galaxies, and in particular the $\rm
	M_{BH}-\sigma$ and the $\rm M_{BH}-M_{spheroid}$ relations.

	\item Stars formed in galactic outflows and escaping the galaxy
	would produce SNe exploding outside galaxies, hence accounting for the
	population of hostless
	SNe observed in galaxy cluster. SNe exploding on large orbits, or outside
	galaxies, can contribute to the {\it in-situ}
	enrichment of the Circum-Galactic Medium and of the Inter-Galactic Medium.
	Such SNe can also result in a very effective heating of the
	galactic halo, hence preventing cold gas accretion onto the galaxy and
	therefore contributing to the delayed quenching of star formation
	in the galaxy as a consequence of starvation.

	\item Young stars formed in the outflow quickly decouple from their
	molecular cloud and therefore are expected to have large escape fraction
	of ionizing photons. As a consequence, if star formation in outflows is common
	to primeval galaxies, this phenomenon can contribute significantly
	to the reionization of the Universe.

	\end{itemize}

\end{itemize}

\section*{Acknowledgments}

The authors are grateful to the anonymous referee for their very useful
comments and suggestions, which really improved the manuscript.
The authors are also grateful to Andy Fabian, Stijn Wuyts, Alessandro Marconi,
Tiago Costa and Kastytis Zubovas for their useful comments. 
R.G. and R.M. acknowledge ERC Advanced Grant 695671 "QUENCH" and support by the Science and Technology Facilities Council (STFC). RR thanks to CNPq and FAPERGS for partial funding this project. Funding for the Sloan Digital Sky Survey IV has been provided by the Alfred P. Sloan Foundation, the U.S. Department of Energy Office of Science, and the Participating Institutions. SDSS acknowledges support and resources from the Center for High-Performance Computing at the University of Utah. The SDSS web site is www.sdss.org.

\setlength{\labelwidth}{0pt}
\bibliographystyle{mn2e}
\bibliography{bibl3}





\appendix
\section{Shock models and diagnostic diagrams}

\begin{figure*}
\centerline{\includegraphics[width=11truecm,angle=90]{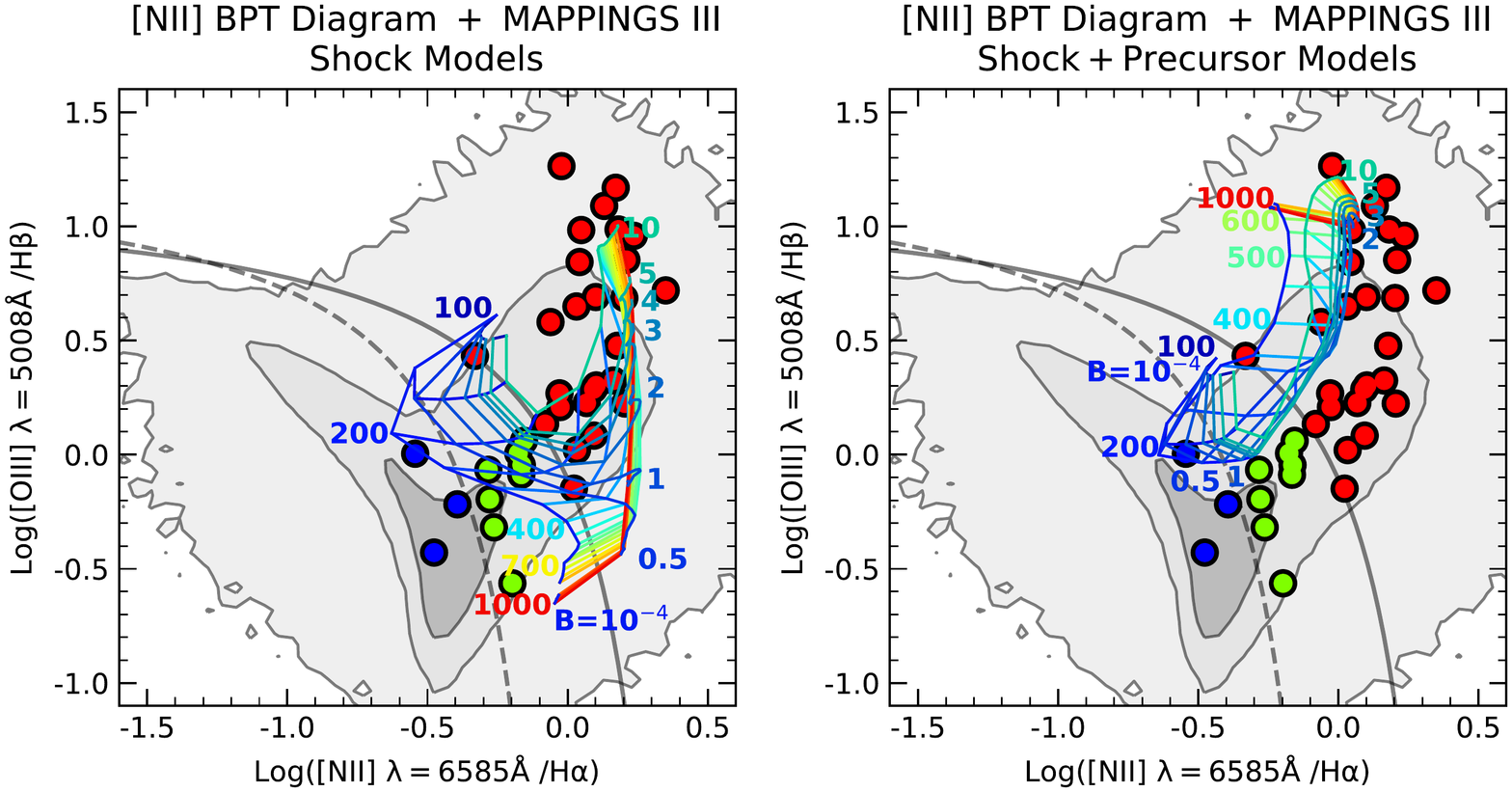}}
\caption{
[NII]-BPT diagram showing the median location of the outflowing gas in each of the galaxies our our sample,
as Fig.~\ref{fig:bpt_total}, but also showing the result of a grid shock models with varying velocities, from 100 km/s (blue) to 1,000 km/s (red),
and varying magnetic field parameter, from $\rm B/n^{1/2}=10^{-4}~\mu G~cm^{3/2}$ (dark blue) to $\rm B/n^{1/2}=10~\mu G~cm^{3/2}$ (light blue),
both wihtout (left) and with (right) shock precursor. Outflows classified as star forming can hardly be accounted for through shock models.
}
\label{fig:shock_NII_BPT}
\end{figure*}

As mentioned in sect.~\ref{sec:shocks}, we have employed MAPPINGS III library of fast radiative shock models
to investigate whether the line ratios observed in star forming outflows could be potentially explained by some form of shocks excitation.
We have used the grid of shock models obtained by \cite{Allen2008}, which span a broad range  of velocities ($\mathrm{v_s=100-1000km/s}$)
and magnetic parameters 
($\mathrm{B/\sqrt{n} = 10^{-4}-10 \mu G cm^{3/2}}$). We have adopted  solar abundances \citep[appropriate for the central region of these galaxies,
][]{Belfiore2017c} and 
pre-shock densities of $\mathrm{1cm^{-3}}$ (which is the case most extensively studied by Allen et al. 2008 and generally regarded as appropriate
for the pre-shock conditions). The resulting grids of models are plotted on the three BPT diagrams and on the R$_{23}$ vs O$_{32}$ diagram
in Figs. \ref{fig:shock_NII_BPT} to \ref{fig:u},
where we also overplot the line ratios observed in the Manga outflows. As already mentioned in sect.~\ref{sec:shocks}, while shock models can
reproduce LIER and AGN-like line ratios, the vast majority of outflows that we have identified as ``star-forming'' (blue symbols) are not consistent
with the shock models.

\begin{figure*}
\centerline{\includegraphics[width=11truecm,angle=90]{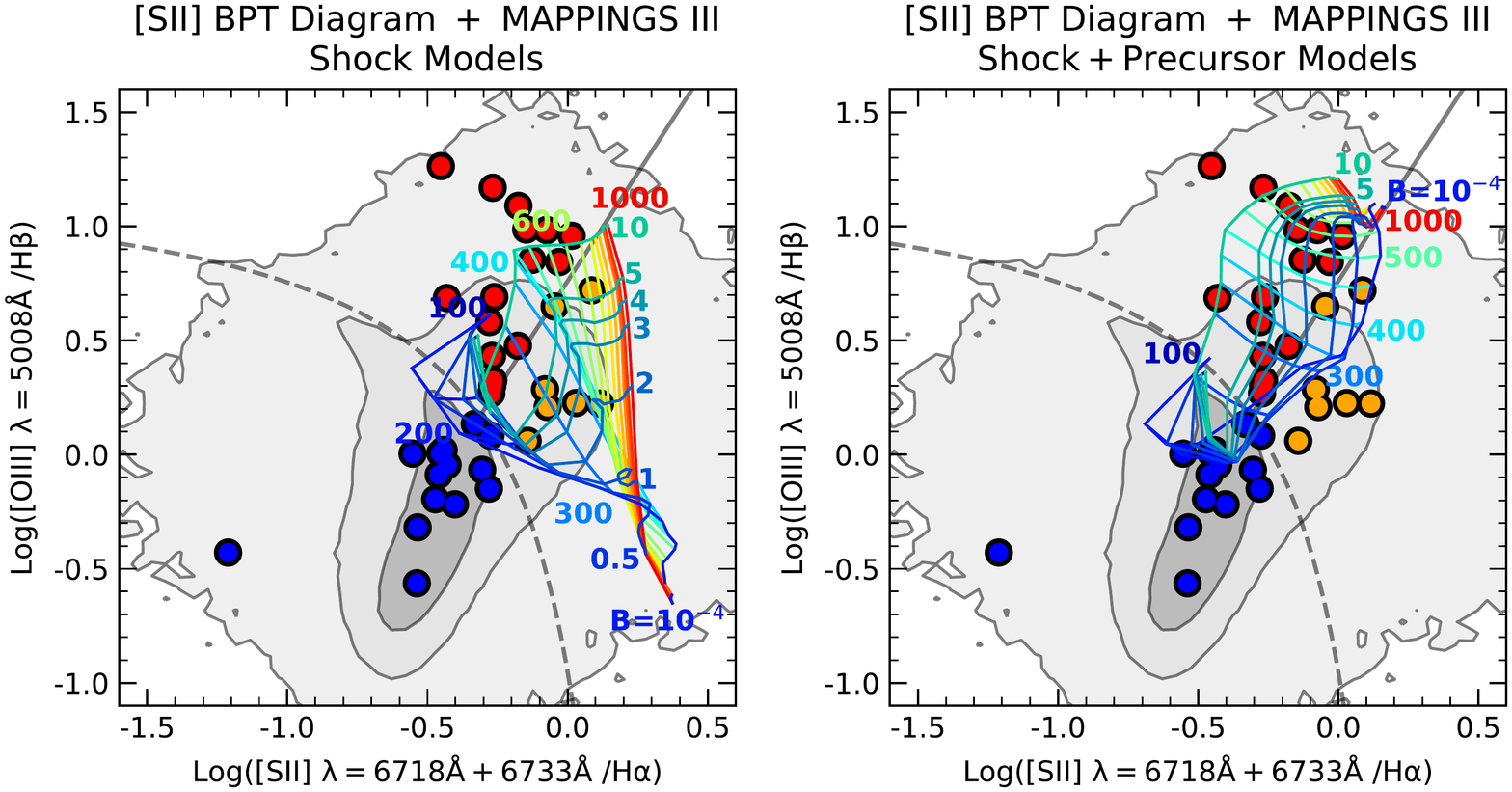}}
\caption{
[SII]-BPT diagram showing the median location of the outflowing gas in each of the galaxies our our sample,
as Fig.~\ref{fig:bpt_total}, but also showing the result of a grid shock models with varying velocities, from 100 km/s (blue) to 1,000 km/s (red),
and varying magnetic field parameter, from $\rm B/n^{1/2}=10^{-4}~\mu G~cm^{3/2}$ (dark blue) to $\rm B/n^{1/2}=10~\mu G~cm^{3/2}$ (light blue),
both wihtout (left) and with (right) shock precursor. Most outflows classified as star forming cannot be explained in terms of shock.
}
\label{fig:shock_SII_BPT}
\end{figure*}

\begin{figure*}
\centerline{\includegraphics[width=11truecm,angle=90]{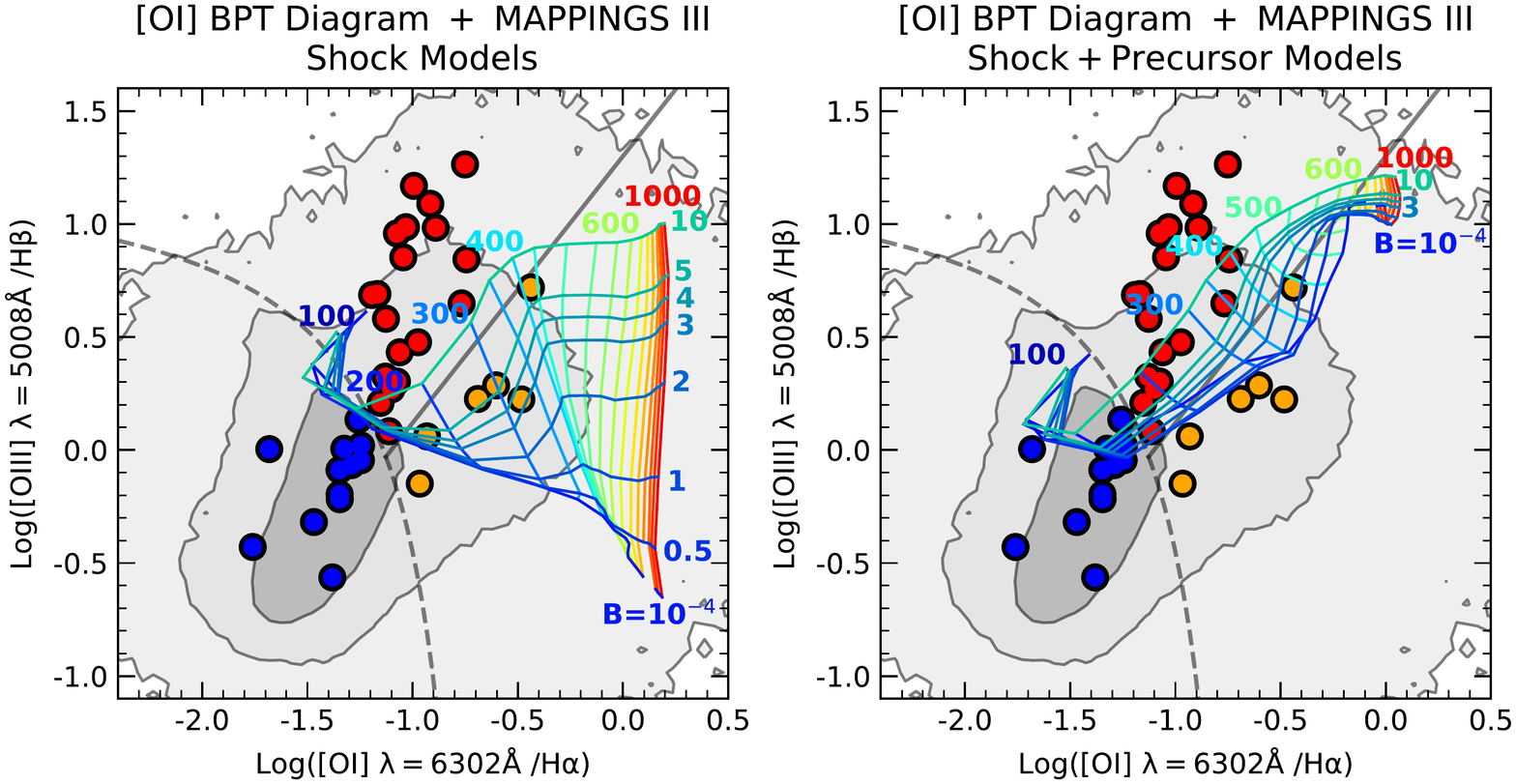}}
\caption{
[OI]-BPT diagram showing the median location of the outflowing gas in each of the galaxies our our sample,
as Fig.~\ref{fig:bpt_total}, but also showing the result of a grid shock models with varying velocities, from 100 km/s (blue) to 1,000 km/s (red),
and varying magnetic field parameter, from $\rm B/n^{1/2}=10^{-4}~\mu G~cm^{3/2}$ (dark blue) to $\rm B/n^{1/2}=10~\mu G~cm^{3/2}$ (light blue),
both wihtout (left) and with (right) shock precursor. Most outflows classified as star forming cannot be explained in terms of shock.
}
\label{fig:shock_OI_BPT}
\end{figure*}

\begin{figure*}
\centerline{\includegraphics[width=11truecm,angle=90]{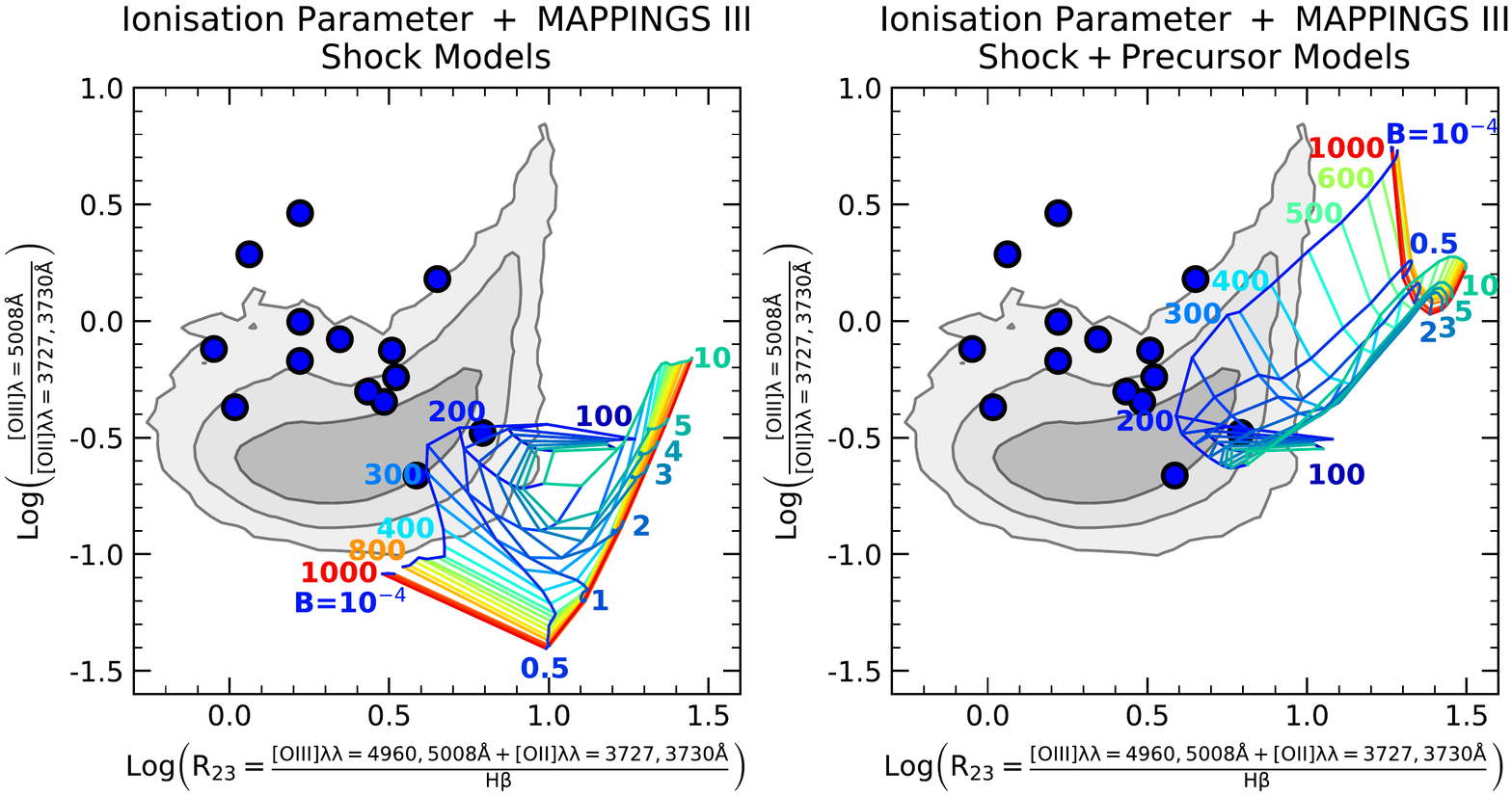}}
\caption{
$\rm R_{23}$ vs [OIII]/[OII] diagram showing the median location of the outflowing gas in each of the galaxies our our sample,
as Fig.~\ref{fig:u}, but also showing the result of a grid shock models with varying velocities, from 100 km/s (blue) to 1,000 km/s (red),
and varying magnetic field parameter, from $\rm B/n^{1/2}=10^{-4}~\mu G~cm^{3/2}$ (dark blue) to $\rm B/n^{1/2}=10~\mu G~cm^{3/2}$ (light blue),
both wihtout (left) and with (right) shock precursor. Most outflows classified as star forming cannot be explained in terms of shock.
}
\label{fig:shock_U}
\end{figure*}

\begin{table*}
\begin{tabular}{|r|r|r|r|c|r|}
\hline
\hline
PID  & IFU   &	RA	     &  DEC	    &  z      & $\rm \log{(M_{star})}$ \\
     &       &  [deg]    &  [deg]   &         & $\rm [M_{\odot}]$      \\
\hline
7443 & 12703 & 229.52558 & 42.74584 & 0.04027 & 10.82$^*$ \\
8241 &  6102 & 126.05963 & 17.33195 & 0.03725 & 10.45  \\
8244 &  3702 & 131.81500 & 51.24580 & 0.02754 & 10.01  \\
8252 &  1902 & 146.09183 & 47.45985 & 0.02589 &  9.35  \\
8256 & 12704 & 166.12940 & 42.62455 & 0.12610 & 11.16$^*$  \\
8318 &  6102 & 197.23931 & 45.90544 & 0.12907 & 11.34  \\
8329 &  3701 & 213.43219 & 43.66248 & 0.08934 &  --$^*$    \\
8341 & 12704 & 189.21325 & 45.65117 & 0.03034 & 10.62  \\
8439 &  6102 & 142.77816 & 49.07974 & 0.03393 & 10.72  \\
8482 & 12704 & 243.58182 & 50.46561 & 0.06025 & 11.39  \\
8550 &  3704 & 248.42638 & 39.18511 & 0.02984 & 10.44  \\
8550 &  9102 & 247.20905 & 39.83508 & 0.03585 & 10.67  \\
8588 &  6101 & 248.45675 & 39.26320 & 0.03176 & 10.74  \\
8612 & 12704 & 254.56457 & 39.39146 & 0.03431 & 10.76  \\
8715 &  3702 & 119.92067 & 50.83997 & 0.05436 & 10.33  \\
8931 &  9101 & 193.07383 & 27.08555 & 0.02098 & 10.25  \\
8946 &  3701 & 168.95772 & 46.31956 & 0.05328 & 10.67  \\
8948 & 12704 & 167.30601 & 49.51943 & 0.07242 & 11.02  \\
8952 &  3703 & 205.44092 & 27.10634 & 0.02878 & 10.33  \\
8979 &  6102 & 241.82338 & 41.40360 & 0.03463 & 10.76  \\
9026 &  9101 & 249.31841 & 44.41822 & 0.03141 & 10.82  \\
9049 &  1901 & 247.56097 & 26.20647 & 0.13145 & 11.38  \\
10001 & 6102 & 132.65399 & 57.35966 & 0.02610 & 10.67  \\
8250 & 12704 & 138.98137 & 44.33276 & 0.03978 & 10.95$^*$  \\
8263 &  3701 & 184.76431 & 46.10679 & 0.03856 & 10.09  \\
8257 & 12701 & 165.49581 & 45.22802 & 0.01999 & 10.38 \\
8464 &  3702 & 187.0635 & 44.45313 & 0.02289 &  --$^*$   \\
7992 &  6104 & 255.27948 & 64.67687 & 0.02713 & 10.21 \\
8078 &  6104 & 42.739429 & 0.369408 & 0.04421 & 10.05 \\
8082 & 12701 & 48.896456 & -1.01628 & 0.02680 & 10.44 \\
8249 &  3703 & 139.72046 & 45.72778 & 0.02643 &  9.79 \\
8250 &  3703 & 139.73996 & 43.50057 & 0.04005 &  9.62 \\
8262 &  9102 & 184.55356 & 44.17324 & 0.02452 & 10.29 \\
8315 & 12705 & 235.92048 & 39.54035 & 0.06347 & 10.60 \\
8550 & 12703 & 247.67443 & 40.52938 & 0.02981 & 10.55 \\
8726 & 12701 & 115.71703 & 22.11273 & 0.02864 & 10.72 \\
8931 & 12702 & 192.76494 & 27.36996 & 0.02772 & 10.07 \\
\hline 
\hline
\end{tabular}
\caption{List of MaNGA galaxies whose outflow can be characterized through the BPT diagrams. The columns provide
the following information: 1) Plate ID; 2) IFU ID; 3) Righ Ascension; 4) Declination; 5) Stella Mass taken from
the MPA-JHU catalog \citep{Brinchmann2004}
(lack of entry means that the stellar mass is not available in the MPA-JHU catalogue).
$^*$Interacting systems.}
\label{tab:z_mass}
\end{table*}

\begin{table*}
\begin{tabular}{|r|r|r|r|r|r|c|c|c|c|c|c|}
\hline
\hline
PID  & IFU   &	 SFR(disk) & SFR(outflow) & $\rm \dot{M}_{outf}(Ha)$ & $\rm \dot{M}_{outf}([OIII])$ &\multicolumn{3}{c}{BPT Class. Disc} & \multicolumn{3}{c}{BPT Class. Outf} \\
     &       &  $\rm [M_{\odot}~yr^{-1}]$& $\rm [M_{\odot}~yr^{-1}]$& $\rm [M_{\odot}~yr^{-1}]$& $\rm [M_{\odot}~yr^{-1}]$    & [NII] & [SII] & [OI] & [NII] & [SII] & [OI] \\
\hline \\
7443 & 12703 &  36.38 &  --  &2.608 & 0.057 & Comp & SF  & LIER & AGN  & AGN  & AGN \\
8241 &  6102 &   5.21 &  --  & 0.37 & 0.036 & Comp & SF  & SF   & AGN  & AGN  & AGN \\
8244 &  3702 &   1.39 &  --  & 0.09 & 0.004 & SF   & SF  & SF   & AGN  & AGN  & AGN \\
8252 &  1902 &   0.07 &  --  & 0.01 & 0.001 & SF   & SF  & LIER & AGN  & AGN  & AGN \\
8256 & 12704 &   2.45 &  --  & 0.59 & 0.038 & AGN  & LIER& AGN  & AGN  & AGN  & AGN \\
8318 &  6102 &   9.13 &  --  & 1.99 & 0.194 & Comp & SF  & AGN  & AGN  & LIER & LIER \\
8329 &  3701 &   0.85 &  --  & 0.15 & 0.015 & Comp & SF  & LIER & AGN  & LIER & LIER \\
8341 & 12704 &   2.45 &  --  & 0.21 & 0.045 & SF   & SF  & SF   & AGN  & AGN  & AGN \\
8439 &  6102 &   4.98 &  --  & 0.61 & 0.023 & Comp & SF  & SF   & AGN  & AGN  & AGN \\
8482 & 12704 &   1.22 &  --  &$<$0.1& 0.014 & AGN  & LIER& AGN  & AGN  & LIER & LIER \\
8550 &  3704 &   0.19 &  --  & 0.34 & 0.012 & AGN  & LIER& SF   & AGN  & LIER & LIER \\
8550 &  9102 &   4.90 &  --  & 0.07 & 0.004 & SF   & SF  & SF   & AGN  & AGN  & AGN \\
8588 &  6101 &   3.82 &  --  & 0.24 & 0.013 & Comp & SF  & SF   & AGN  & LIER & AGN \\
8612 & 12704 &   1.76 &  --  & 0.16 & 0.015 & AGN  & AGN & AGN  & AGN  & AGN  & AGN \\
8715 &  3702 &   2.70 &  --  &16.90 & 0.887 & AGN  & AGN & AGN  & AGN  & AGN  & AGN \\
8931 &  9101 &   0.36 &  --  & 0.15 & 0.001 & Comp & SF  & LIER & AGN  & AGN  & AGN \\
8946 &  3701 &   0.51 &  --  & 0.06 & 0.005 & Comp & SF  & SF   & AGN  & AGN  & AGN \\
8948 & 12704 &   4.43 &  --  & 0.17 & 0.013 & Comp & SF  & SF   & AGN  & AGN  & AGN \\
8952 &  3703 &   0.24 &  --  & 0.02 & 0.002 & Comp & SF  & SF   & SF   & LIER & LIER \\
8979 &  6102 &   1.37 &  --  & 0.05 & 0.004 & SF   & SF  & SF   & AGN  & LIER & AGN \\
9026 &  9101 &   1.23 &  --  & 0.20 & 0.008 & Comp & SF  & SF   & AGN  & AGN  & AGN \\
9049 &  1901 &  15.24 &  --  & 0.49 & 0.033 & Comp & SF  & SF   & AGN  & AGN  & AGN \\
10001 & 6102 &   1.17 &  --  & 0.24 & 0.004 & Comp & SF  & SF   & AGN  & AGN  & AGN \\
8250 & 12704 &  16.57 & 4.30 & 0.33 & 0.009 & Comp & SF  & SF   & AGN  & SF   & AGN \\
8263 &  3701 &   0.61 & 0.31 & 0.13 & 0.003 & Comp & SF  & SF   & AGN  & SF   & LIER \\
8257 & 12701 &  5.824 & 0.64 & 0.22 & 0.016 & Comp & SF  & SF   & AGN  & SF   & SF \\
8464 &  3702 &   2.20 & 0.17 & 0.12 & 0.010 & Comp & SF  & SF   & AGN  & SF   & SF \\
7992 &  6104 &   2.86 & 0.29 & 0.09 & 0.003 & Comp & SF  & SF   & Comp & SF   & SF \\
8082 & 12701 &   1.10 & 0.09 & 0.06 & 0.004 & SF   & SF  & SF   & Comp & SF   & SF \\
8262 &  9102 &   5.00 & 0.93 & 0.36 & 0.051 & SF   & SF  & SF   & Comp & SF   & SF \\
8315 & 12705 &   5.74 & 1.04 & 0.26 & 0.021 & SF   & SF  & SF   & Comp & SF   & SF \\
8550 & 12703 &   1.11 & 0.09 & 0.04 & 0.004 & SF   & SF  & SF   & Comp & SF   & SF \\
8726 & 12701 &   0.93 & 0.23 & 0.11 & 0.009 & Comp & SF  & SF   & Comp & SF   & SF \\
8931 & 12702 &   1.72 & 0.10 & 0.05 & 0.003 & SF   & SF  & SF   & Comp & SF   & SF \\
8078 &  6104 &   1.32 & 0.88 & 0.24 & 0.027 & Comp & SF  & SF   & SF   & SF   & SF \\
8249 &  3703 &   0.65 & 0.03 & 0.01 & 0.002 & SF   & SF  & SF   & SF   & SF   & SF \\
8250 &  3703 &   1.42 & 0.29 & 0.07 & 0.016 & SF   & SF  & SF   & SF   & SF   & SF \\
\hline
\hline
\end{tabular}
\caption{Properties of the star formation rates, outflow rate and BPT classification of the discs and of
the outflows. The columns provide the following information:
1) Plate ID; 2) IFU ID; 3) Star Formation Rate in the disc; 4) Star Formation Rate in the outflow (only for those
galaxies that are dominated by star formation in the outflow according to the [SII]-BPT diagram); 5) Ionized outflow rate
determined through the broad H$\alpha$ emission; 6) Ionized outflow rate
determined through the broad [OIII] emission; 7--9) BPT classification of the nebular emission of the galaxy disc;
10--12) BPT classification of the nebular emission of the outflow.
}
\label{tab:sfr_outf}
\end{table*}

\end{document}